\documentclass[journal=langd5,manuscript=article]{achemso}
\usepackage[version=3]{mhchem} 
\setkeys{acs}{usetitle = true}

\author{Kai Huang}
\affiliation
{Materials Science Program, University of Wisconsin, Madison, Wisconsin 53706-1595, USA}
\author{Izabela Szlufarska}
\email{szlufarska@wisc.edu}
\affiliation
{Department of Materials Science \& Engineering, University of Wisconsin, Madison, Wisconsin 53706-1595, USA}
\title[\texttt{achemso} demonstration]
{Friction and slip at solid/liquid interface in vibrational systems}
\begin{document}
\singlespacing
\begin{abstract}
Molecular dynamics simulations have been performed to study frictional slip and its influence on energy dissipation and momentum transfer at atomically smooth solid/water interfaces. By modifying surface chemistry, we investigate the relationship between slip and the mechanical response of a vibrating solid for both hydrophilic and hydrophobic surfaces. We discover physical phenomena that emerge at high frequencies and that have significant contributions to energy dissipation. A new analytical model is developed to describe mechanical response of the resonators in this high frequency regime, which is relevant in such applications as MEMS-based biosensors. We find a linear relationship between the slip length and the ratio of the damping rate shift to resonant frequency shift, which provides a new way to obtain information about slip length from experiments.
\end{abstract}
\section{Introduction}
\begin{figure}
\includegraphics[scale=0.6]{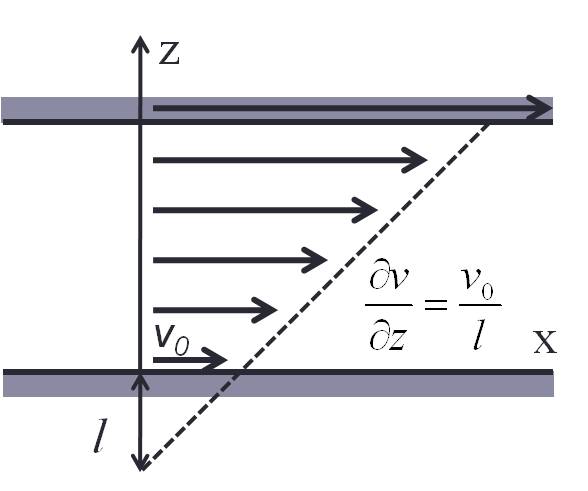}
\caption{Schematic representation of the slip boundary condition with slip length $l$. Here, the velocity of the solid $u_0$=0.}
\label{f.definition}
\end{figure}

Friction at solid/liquid interfaces plays an important role in many mechanical devices. An example is quartz crystal microbalance (QCM)~\cite{QCM-review,vapor,liquid-overlayers,application3,application2,application}, which in recent years has become a widely used mechanical method for characterization of bio-interfaces. QCM provides also a direct experimental approach to study friction~\cite{QCM-friction1, QCM-friction2, mark} since its acoustic shear-wave motion is sensitive to the sliding friction at its surface. Such interfacial friction will result in a shift of the resonant frequency $f_0$ and in the case quartz crystal with dissipation monitoring (QCM-D), also in a shift of a damping rate $D$. $f_0$ and $D$ are defined as
\begin{equation}
f_0=\frac{1}{2d}\sqrt{\frac{c_{q}}{\rho_{q}}}
\label{e.frequency}
\end{equation}
\begin{equation}
D=\frac{-\Delta E_n}{2E_n},
\label{e.damping}
\end{equation}
where $d$, $c_{q}$ and $\rho_{q}$ are the thickness, stiffness, and density of quartz, respectively, and $E_n$ is the mechanical energy stored in quartz during the $n^{\mathrm{th}}$ vibrational cycle.
Solid/liquid interface is viscous in nature and the corresponding friction force can be written as
\begin{equation}
F=-\overline{\eta}(v_0-u_0),
\label{e.viscous}
\end{equation}
where $\overline{\eta}$ is the friction coefficient and $v_0$ and $u_0$ are the shear velocity of liquid and solid at the interface, respectively. The term $v_0-u_0$ is the slip velocity, defined as the sliding velocity of the liquid adjacent to the solid relatively to the solid surface. For sufficiently large friction coefficients, the slip velocity becomes negligible, which corresponds to the no-slip boundary condition. With no-slip boundary condition, the mechanical response of QCM can be predicted by solving continuum-level wave equations, without the need to know the value of the friction coefficient. However, the assumption of no-slip boundary condition does not always hold and therefore there is a need to develop theories that will take the existence of slip directly into account. The existence of slip has been first proposed over a century ago by Navier~\cite{slip-definition}, but it has been accepted only in recent years~\cite{limits-no-slip, molecular-shape,slip-evidence,AFM}. Slip can be quantified either using the slip velocity $v_0-u_0$ or the slip length $l$ (see~Fig.~\ref{f.definition}), where the latter quantity is defined as
\begin{equation}
l=(v_0-u_0)\left(\frac{\partial v}{\partial z}\right)^{-1}.
\label{e.definition}
\end{equation}
The slip length is proportional to the viscosity $\eta$ of liquid and to the inverse of the friction coefficient $\overline{\eta}$, that is 
\begin{equation}
l=\frac{\eta}{\overline{\eta}}.
\label{e.definition_mechanical} 
\end{equation}

It is now accepted that the slip length can span a wide range of values, from several Angstroms (a molecular diameter scale) to micrometers for super-hydrophobic surfaces~\cite{superhydrophobic}, and that contributions from slip to dynamics at the solid/liquid interface cannot be neglected.
Large slip length is likely to occur when hydrophilic surfaces meet hydrophobic liquids or vice versa, both scenarios frequently encountered in biological systems. Slip is also expected to play an important role in resonators with high resonance frequency in the region of upper MHz ($\sim$100 MHz) and possibly even GHz. This is because at such high frequencies the penetration length of liquid is on the micron scale and generally slip is more important when its length scale becomes comparable to the size of the system of interest. Another example of application where slip plays an important role is the microfluidics~\cite{microdevices,future-of-microfluidics,microfluidics,high-frequency-nanofluidics}. In this case a tiny amount of liquid flows through nano-/micro-scale pipes and friction at the pipe wall can significantly affect the flow due to the high surface/volume ratio.

Not surprisingly, understanding of the solid/liquid slip has been argued to be an important challenge in research on solid/liquid interfaces.~\cite{slip-review1,slip-review2,slip-review3}
One of the difficulties in investigating slip at solid/liquid interfaces is that the slip length is difficult to measure experimentally. Such measurements require a very high sensitivity of the experimental apparatus to the shear stress of liquid as well as a control of the surface quality, which includes both surface roughness and chemistry. Atomistic simulations and in particular molecular dynamics (MD) simulations, provide a powerful tool to complement experiments and to bring insights into slip-related phenomena. The MD technique enables a precise control of such factors as shear rate and vibrational frequency, and makes it possible to analyze fluid velocity gradient close to the solid/liquid interface. In addition, thanks to the ability to model atomically smooth surfaces, in MD simulations it is possible to isolate effects of surface chemistry (e.g., hydrophobicity) from effects of surface roughness. MD simulations have been already employed to determine the dependence of slip on shear rate, chemical bond strength, and surface roughness~\cite{wetting-slip, nanofluidics, interfacial-water, contact-slip1, contact-slip2, polymer-shear, high-shear-rates, shear2, shear3, Zhang1, Zhang2}. For example, Barrat {\it et al.}~\cite{wetting-slip} found that the slip length of water on diamond-like solid surface scales approximately as an inverse of the square of interfacial bond strength $\varepsilon$ between the liquid and the solid, that is: 
\begin{equation}
l\sim\varepsilon^{-2}.
\label{e.slip_bond}
\end{equation}
In other simulation studies~\cite{Thompson, polymer-shear, shear2}, slip length has been observed to increase with shear rate $\dot{\gamma}$, consistently with the following empirical relation 
\begin{equation}
l(\dot{\gamma})=l_0\left(1-\dot{\gamma}/\dot{\gamma}_c\right)^{-0.5},
\label{e.slip_shear}
\end{equation}
where $l_0$ is the intrinsic slip length, which corresponds to the limit of zero shear rate, and $\dot{\gamma}_c$ is the maximum shear rate the a given liquid can carry.
The aforementioned simulations of slip phenomena typically involve a sliding system in its steady state or, more specifically, with a constant shear rate built in by confining the liquid between two parallel solid walls. The slip length measured in this way is limited to a non-vibrating (here, referred to as static) system and therefore it cannot provide any dynamic (i.e., related to vibrations) information about the solid/liquid friction. In this study we investigate the effect of slip on mechanical properties of a vibrating system, such as QCM, and therefore it is necessary to first extend the concept of a static slip ($l_s$) to the dynamic slip ($l_d$) and to discuss the relationship between these two quantities. The dynamic slip length can generally be a function of both shear rate $\dot{\gamma}$ and frequency $\omega$, and it can be written as $l_d(\dot{\gamma},\omega)$.
In the case of a small amplitude vibration associated with a small shear rate, the dynamic slip length is approximately independent of the shear rate. In this limit, we can simplify the dynamic slip length to be $l_d(\omega)$. Later on, we will focus on the dynamic slip length in the small shear rate limit and the symbol $l_d$ will always refer to this case. Similarly to the behavior of static slip length at low shear rate, the dynamic slip length will also converge to the intrinsic slip length $l_0$, which is the slip length in the limit $\dot{\gamma}\rightarrow 0$, $\omega\rightarrow 0$. The intrinsic slip length $l_0$ depends on the properties of the solid and the liquid and on the interfacial geometry. In our simulations, for simplicity we control the value of $l_0$ by modifying the bond strength $\varepsilon$ between liquid and solid rather than by employing different types of liquid/solid combinations or different interfacial geometries. For mathematical convenience we define
\begin{equation}
\Gamma_s(\varepsilon,\dot{\gamma})= l_s(\varepsilon,\dot{\gamma})/l_0(\varepsilon),
\label{e.ls}
\end{equation}
\begin{equation}
\Gamma_d(\varepsilon,\omega)= l_d(\varepsilon,\omega)/ l_0(\varepsilon),
\label{e.ld}
\end{equation}
$\Gamma_s$ ($\Gamma_d$) is the ratio between the static slip length (the dynamic slip length) and the intrinsic slip length. $\Gamma_s$ ($\Gamma_d$) is expected to be equal to unity when the shear rate (frequency) is zero. 
As we can control the intrinsic slip length $l_0$ by changing the strength of interfacial bonds in simulation, we are able to investigate the influence of the slip length on the momentum transfer and energy dissipation at the solid/liquid interface with a particular focus on the transverse-shear model of a vibrating interface. We focus on a Newtonian liquid and a simple (i.e., unpatterned) solid surface, which is of relevance for most QCM applications. For a Newtonian liquid, the damping wave through the liquid can be well described using a single parameter called penetration length~\cite{Bird}, which is defined as
\begin{equation} 
\delta=\sqrt{\frac{2\eta}{\omega\rho_{l}}},
\label{e.penetration}
\end{equation}
where $\rho_{l}$ is the density of a liquid. The shear velocity amplitude of the damping wave along the $z$ direction (see Fig.~\ref{f.vibrating_system}) can be written as $|v(z)|=|v_{0}|e^{-z/\delta}$, where $v_0=|v_{0}|e^{i\omega t}$ is the velocity of the liquid adjacent to the solid surface. The penetration length $\delta$ describes how fast the shear wave emitted at the vibrating interface decays when it travels through the liquid. The QCM system can only sense the viscosity of liquid within a distance of a few times the penetration length from the QCM surface as the amplitude of damping wave of liquid at larger distances is low and can be ignored. The relationship between $v_0$ and the vibrational velocity $u_0$ of the solid's interface is related to the penetration length through the following equation~\cite{Persson}
\begin{equation} 
v_0=\frac{u_0}{1-(i-1)\frac{l}{\delta}}.
\label{e.bcv}
\end{equation}
From the above expression it is clear that when the slip length $l$ is much smaller than the penetration length $\delta$, then $v_0\approx u_0$, which corresponds to the no-slip boundary condition. If we define a normalized slip length $b=l/\delta$ ($b_0$, $b_s$ and $b_d$ for the intrinsic, static, and dynamic cases, respectively), we can see that boundary slip becomes important when $b$ is not negligible as compared to 1. In our simulation, a wide range of $b$ values can be accessed by varying the interfacial bond strength (which controls surface hydrophobicity) and the vibrational frequency (which controls the penetration length). The ability to achieve this wide span of normalized slip lengths allows us to determine relationship between QCM's mechanical response and slip. Such relationship is necessary to fully understand the effect of slip boundary condition on solid/liquid interfacial momentum transfer and energy dissipation. 
In subsequent sections we first review continuum level theories that are currently used to interpret QCM experiments. We begin with theories that assume no-slip boundary conditions, followed by a discussion of how slip boundary conditions have been introduced into these models. As the existing slip boundary models have not been validated in experiments nor in simulations, we will test these theories using MD simulations. We discuss what physical phenomena are not captured in the existing theories and we provide a new model that includes these phenomena.

\section{Review of continuum-level models}
The first continuum-level theory for QCM came from Sauerbrey~\cite{sauerbrey}, who provided the relation between the frequency shift of QCM and the mass attached to it. The Sauerbrey relation assumes that the attached mass is a thin rigid (i.e., infinitely stiff) film and therefore no energy dissipation takes place in the attached film. According to the Sauerbrey theory, the shift $\Delta f$ of frequency can be related to $\Delta m$, which is the mass of the attached film per unit area, as follows: 
\begin{equation}
\Delta f=\frac{-2f_{0}^2}{\sqrt{c_{q}\rho_{q}}}\Delta m.
\label{e.sauerbrey}
\end{equation}
In this expression \(f_{0}\) stands for the resonant frequency of the unloaded system (without the attached film), and $c_{q}$ and $\rho_{q}$ have the same meaning as in Eq.~\ref{e.frequency}.
When QCM is placed in an aqueous environment, as often required in applications of biosensing, vibrational energy of the QCM is damped into the liquid. This damping occurs as the result of a viscous coupling, or in other words, by transmission of the shear acoustic waves across the solid/liquid interface.
For a Newtonian liquid with non-slip conditions at the solid-liquid interface, one can solve the problem of wave propagation analytically to predict the change ($\Delta D$) in the damping factor and the shift ($\Delta f$) in the natural frequency of the solid due to the presence of the liquid (referred to as liquid-loading). A mathematical formalism for this problem has been first introduced by Kanazawa and Gordon~\cite{kanazawa}, who solved coupled wave-propagation and the Navier-Stokes equations. The resulting solution predicts that the resonance frequency (damping rate ) of the QCM decreases (increases) due to the presence of the liquid with viscosity $\eta$ and density $\rho_{l}$ as: 
\begin{equation}
\Delta f=-f_0^{3/2}\sqrt{\frac{\rho_{l}\eta}{\pi\rho_{q}c_{q}}},
\label{e.kanazawa_f}
\end{equation}
\begin{equation}
\Delta D=-2\pi\frac{\Delta f}{f_0}.
\label{e.kanazawa_d}
\end{equation}
Martin {\it et al.}~\cite{martin} considered the case of a combined loading of a thin rigid film and an infinitely deep Newtonian liquid. The authors proposed that contributions to $\Delta f$ from the film and the liquid are additive. It has been later found that Martin's additive model overestimates $\Delta f$ observed experimentally for the combination of a soft film and a liquid, which phenomenon has been called a "missing mass effect". The missing mass effect has been attributed to the presence of viscous coupling between the soft film and the liquid. A model that takes this physics into account has been developed by Voinova {\it et al.}~\cite{missing-mass} Both, the Martin's and Voinova's models assume no-slip boundary conditions.
Studies that take into account slip boundary conditions explicitly are much scarcer due to the difficulties discussed in the introduction section. Nevertheless there have been several theoretical studies aimed at incorporating the slip effect into the Kanazawa model~\cite{complex-slip, Hayward, Daikhin, Mchale, Zhuang, Persson, bubble-slip}. For example, Ferrante {\it et al.}~\cite{complex-slip} introduced a complex interfacial slip parameter $\alpha$. However, a physical meaning has not been provided for the two fitting parameters (the real and the imaginary parts of $\alpha$) that appear in the model. Other theoretical approaches to slip boundary conditions include a model by Ellis {\it et al.}~\cite{Hayward}, who proposed a relation between the real and imaginary parts of the complex slip parameter and thereby was able to replace the complex slip parameter $\alpha$ with a single parameter, which is the slip length. This new model provided a clear connection between the QCM's response and the slip length and it has been invoked to explain a number of experimental results. For instance, Daikhin {\it et al.}~\cite{Daikhin} applied this model in the studies of adsorption of pyridine on gold surfaces and to explain the observed difference between the prediction of no-slip theory and experimental results. McHale {\it et al.}~\cite{Mchale} used loading impedance to analyze similar discrepancy between experimental observations on rough surfaces and the no-slip boundary model and introduced the concept of a negative slip length to explain the discrepancy. Zhuang {\it et al.}~\cite{Zhuang} followed classic hydrodynamic theories to derive a mathematical formalism for the slip boundary condition and extended their model to the non-Newtonian regime in order to explain the surprising experimental observation that frequency shift $\Delta f$ can be positive (no-slip models do not allow the frequency to increase in a Newtonian liquid). A mathematical analysis of a vibrational interface with slip boundary conditions and with a simplified solid (a spring attached to a solid slab) has been also reported by Persson ~\cite{Persson}. 
It is straightforward to show that ignoring the roughness of the surface and the width of the interface (see discussion section), all the one-parameter slip-boundary models from Refs.~\cite{Hayward,Daikhin, Mchale, Zhuang, Persson} can be reduced to the following set of equations:
\begin{equation}
\frac{\Delta f}{f_0}=-\frac{1}{\pi Z}\sqrt{\frac{\rho_{l}\eta\omega}{2}}\frac{1}{1+2b_0+2b_0^2}
\label{e.0slip_f}
\end{equation}
\begin{equation}
\Delta D=\frac{2}{Z}\sqrt{\frac{\rho_{l}\eta\omega}{2}}\frac{1+2b_0}{1+2b_0+2b_0^2}
\label{e.0slip_d}
\end{equation}
where $\omega=2\pi f$ is the angular frequency and $Z=\sqrt{c_{q}\rho_{q}}$ is the mechanical impedance of QCM. $b_0=l_0/\delta$ is the normalized intrinsic slip length we defined in the introduction section. In the limit of $b_0\rightarrow 0$, Eq.~\ref{e.0slip_f} and Eq.~\ref{e.0slip_d} are reduced to the corresponding expressions in the Kanazawa model~\cite{kanazawa}. 

\section{Simulation setup}
All MD simulations have been performed using the LAMMPS software package~\cite{lammps}. 
In our simulations we choose water as a model liquid. We use the TIP4P model~\cite{TIP4P} for water interactions because it correctly describes mechanical properties of water, such as viscosity. Long-range electrostatic interactions are calculated using the Particle-Particle Particle-Mesh method (PPPM) method~\cite{PPPM}. The non-bonded interactions involving hydrogen are not considered and an additional constraining force is applied to hydrogen atoms with the SHAKE~\cite{shake} algorithm to enable a simulation time step as large as 4.0 fs.
Because we are interested in the effects of liquid on the vibrational properties of a solid instead of materials properties of the solid itself, we choose a model solid in which atoms interact via Lenard-Jones (LJ) force field:
\begin{equation}
U(r)=4\varepsilon \left[\left(\frac{\sigma}{r}\right)^{12}-\left(\frac{\sigma}{r}\right)^{6}\right].
\end{equation}

\begin{table}
\caption{Parameters of the Lennard-Jones potential. * symbol refers to all atom types other than H}
\label{tbl:example}
\begin{tabular}{lll}
\hline
Atom types & $\varepsilon$ (kcal/mol) & $\sigma$(\AA)\\
\hline
H-* & 0.0 & 0.0\\ 
O-O & 0.16275 & 3.16435\\
O-solid & 0.05-1.0 & 2.8\\
solid-solid & 8.0 & 3.368\\
\hline
\end{tabular}
\label{t.lj}
\end{table}

Parameters $\varepsilon$ and $\sigma$ for the LJ force field used in our simulations can be found in Tab.~\ref{t.lj}. The cut-off for interactions is taken to be 12~\AA. 
The solid has face-centered-cubic (fcc) structure with the (100) surface being in contact with the liquid. The value of $\varepsilon$ = 8.0 kcal/mol is sufficient to make the solid sufficiently rigid in our simulations. The properties of the solid wall could affect the slip at the solid/liquid interface~\cite{wall, high-shear-rates}. We choose the flexible wall model (solid atoms are allowed to vibrate) over the rigid wall model (solid atoms are held at their lattice sites). However, in our case the difference in the slip between the two wall models, if any such difference exists, is not expected to be large, since the atomic mass of our solid is about 2 orders of magnitude larger than the molecular mass of water and therefore the vibrational amplitude of atoms in the solid is significantly smaller than of the liquid molecules. Heavy solid atoms are required to make the resonator in our simulation computationally inexpensive and stable during high frequency vibration. The control of interfacial bond strength is realized by choosing the value of $\varepsilon$ for the oxygen-solid interaction. For no-slip boundary condition simulations, the value of $\varepsilon$ is chosen to be 1.0 kcal/mol, which is strong enough to eliminate the slip velocity between solid and liquid. For slip boundary condition simulations, the value of $\varepsilon$ changes from 0.05 to 0.6 kcal/mol, which presents a wide range of slip lengths that are typically found in simple solid/liquid interfaces. We choose the x axis to coincide with the direction of shear velocity and the gradient of velocity to lie along the z axis. Nose-Hover thermostat is coupled to the y and z components of the velocity so that the vibrational mechanical energy in x direction is not affected by the thermostat. We have confirmed that this method of temperature control leads to an exponential decay of the amplitude of free oscillations with time, as expected from theory. We have also performed simulations without any thermostat in the liquid (thermostat is only coupled to atoms in the solid), similarly as was done in MD simulations of the Couette flow reported in Refs.~\cite{arun1,arun2}. We found that the main effect of removing the thermostat from the liquid region is a decrease in water viscosity by about $10\%$ and that this procedure does not affect the slip length and the slip model we propose in subsequent sections. For consistency, all results presented in this paper have been obtained using 2D thermostat applied to both the solid and the liquid regions. In our simulations the system is first relaxed at 300K and 1 atmosphere using NPT ensemble with coupling constants $\tau_{T}$=100 fs for temperature and $\tau_{p}$=1000 fs for pressure. Simulations of QCM vibrations are performed in NVT ensemble.

\begin{figure}
\includegraphics[scale=0.4]{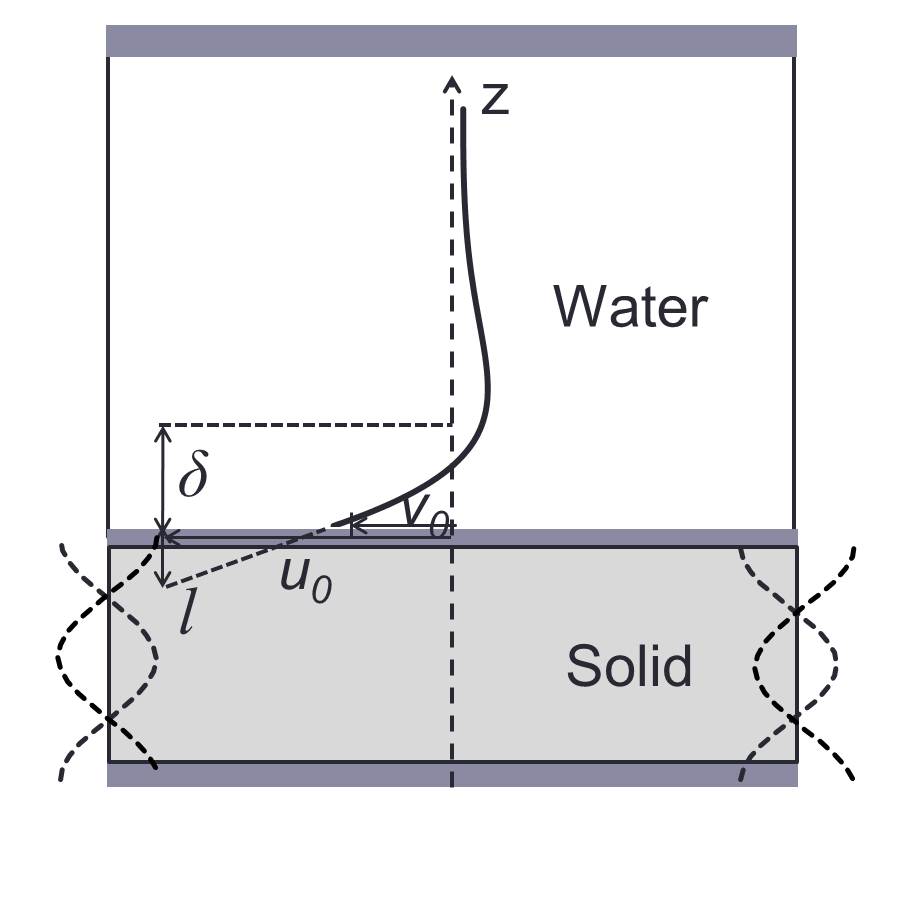}
\caption{Illustration of the acoustic shear wave simulation system with slip length \(l\). \(u_0\) and \(v_0\) are the velocities of solid and liquid at the interface, respectively. \(\delta\) is the penetration length of water, which characterizes the damping wave. }
\label{f.vibrating_system}
\end{figure}

We perform three types of MD simulations. Measurement of water viscosity is carried out using the Reversed Non-Equilibrium Molecular Dynamics (RNEMD) method~\cite{RNEMD1,RNEMD2}. The RNEMD method constrains the velocity of water molecules to achieve a steady state shear rate through the thickness of the liquid. The second simulation type involves a static shearing system to measure the static slip length directly based on Eq.~\ref{e.definition}. In this setup two parallel solid slabs slide with respect to each other to produce a velocity gradient through the liquid confined between the two slabs. Each of the two solid slabs has dimension of 32\AA$\times$32\AA$\times$8\AA~(288 atoms) and the slabs are placed in a simulation box with dimensions of 32\AA$\times$32\AA$\times$81\AA. There are 2000 water molecules placed between the solids, which corresponds to the water thickness of 63\AA. This thickness is large enough to avoid nano-scale confinement effects reported in literature~\cite{confinement}. The third type of simulation involves modeling a vibrational system and measuring its mechanical response to applied force, such as frequency shift and damping rate shift. In this system a thicker box of water is placed above a vibrating solid. One-atom thick solid wall is placed above the water to prevent it from evaporating. The water region has dimensions of 32\AA$\times$32\AA$\times$251\AA~and contains 8000 water molecules. The corresponding water density is 0.99 g/cm$^3$.
Furthermore, we have considered two types of vibrational systems. The first one is a shear-wave QCM resonator (see Fig.~\ref{f.vibrating_system}). We impose acoustic shear wave by initially deforming the resonator using a cosine wave function through the thickness of the QCM (along the z direction) and then removing the constraint and allowing free oscillation of the system. This acoustic shear wave in the QCM resonator forms a standing wave while in the liquid it becomes a damping wave with the source at the solid/liquid interface. The shear modulus of the solid crystal in our simulation is 32.2 GPa. With this system setup we can only study vibrational frequencies above 30 GHz. In order to study a wider frequency range, we simplify the QCM model to a spring model, in which a solid slab is attached to a spring (each atom is constrained with a spring force that pulls it to its initial position), since it enables a shorter computational time for the same vibrational period. Although the properties of water above the vibrating solids are the same for the two types of motions, the mathematical descriptions of the mechanical response of the two resonators are different. In Tab.~\ref{t.simplified}, we provide expressions for the equivalent properties in the shear wave and the spring models, which properties include impendence $Z$, resonant frequency $f_0$, relative change in the frequency $\Delta f/f_0$ in the presence of liquid, and damping shift $\Delta D$ due to the presence of liquid. 

\begin{table}
\caption{Mechanical properties of the two types of resonators discussed in the text. $Z$ is impendence (units of N$\cdot$~kg$\cdot$~m$^{-5})^{0.5}$), $f_0$ is resonant frequency, $\Delta f$ is frequency shift, $\Delta D$ is damping rate, $c_q$ is shear modulus of quartz, $\rho_q$ is density of quartz, $\eta_l$ is viscosity of liquid, $\rho_l$ is density of liquid, $k$ is spring constant per unit, $M$ is mass of solid per unit, and $\omega$ is angular frequency}
\label{tbl:example}
\begin{tabular}{ll}
\hline
Shear wave model & Spring model\\
\hline
\(Z=\sqrt{c_{q}\rho_{q}}\) & \(Z=\sqrt{kM}\)\\
\(f_0=\frac{1}{2d}\sqrt{\frac{c_{q}}{\rho_{q}}}\) & \(f_0=\frac{1}{2\pi}\sqrt{\frac{k}{M}}\)\\
\(\frac{\Delta f}{f_0}=-\frac{1}{\pi Z}\sqrt{\frac{\rho_{l}\eta_{l}\omega}{2}}\) & \(\frac{\Delta f}{f_0}=-\frac{1}{2Z}\sqrt{\frac{\rho_{l}\eta_{l}\omega}{2}}\)\\
\(\Delta D=-2\pi\frac{\Delta f}{f_0}\) & \(\Delta D=-2\pi\frac{\Delta f}{f_0}\)\\
\hline
\end{tabular}
\label{t.simplified}
\end{table}

In the spring model, we control the spring constant $k$ and atomic mass $M$ per unit area, and thereby we vary the resonant frequency while keeping the mechanical impedance \textit{Z} constant and equal to $4.93 \times 10^7$ (N$\cdot$~kg$\cdot$~m$^{-5})^{0.5}$. The frequency in our simulations varies from 4.07 GHz to 65.1 GHz (the shortest period about 15 ps is still 3 orders of magnitude longer than the 4.0 fs time step). The typical resonant frequency of QCM in experiment is on the order of 10 MHz, and the highest frequency of acoustic shear wave devices can currently reach 1GHz, which is on the same order of magnitude as the lower end of frequency range attained in our simulations. The thickness of water in the vibrating system is 251\AA~and it is much larger than the penetration lengths $\delta$, which is found to range from 15\AA~to 76\AA. The simulation therefore provide a good approximation of an infinitely thick liquid, since the QCM can only sense the liquid within a distance from its surface equal to a few times the penetration length (see the introduction section). The penetration length is measured by computing and analyzing the vibrational amplitude along the direction of wave propagation.
In our study we will first demonstrate that both, the QCM shear-wave model and the spring solid model, capture correctly behavior of resonators in the limit of no-slip. We will then use the spring solid model to investigate and provide insights into the slip behavior.

\section{Results}
\subsection{No-slip interface}

\begin{figure} 
\begin{tabular}{cc}
\includegraphics[width=0.36\textwidth,viewport=0 0 700 540,clip=]{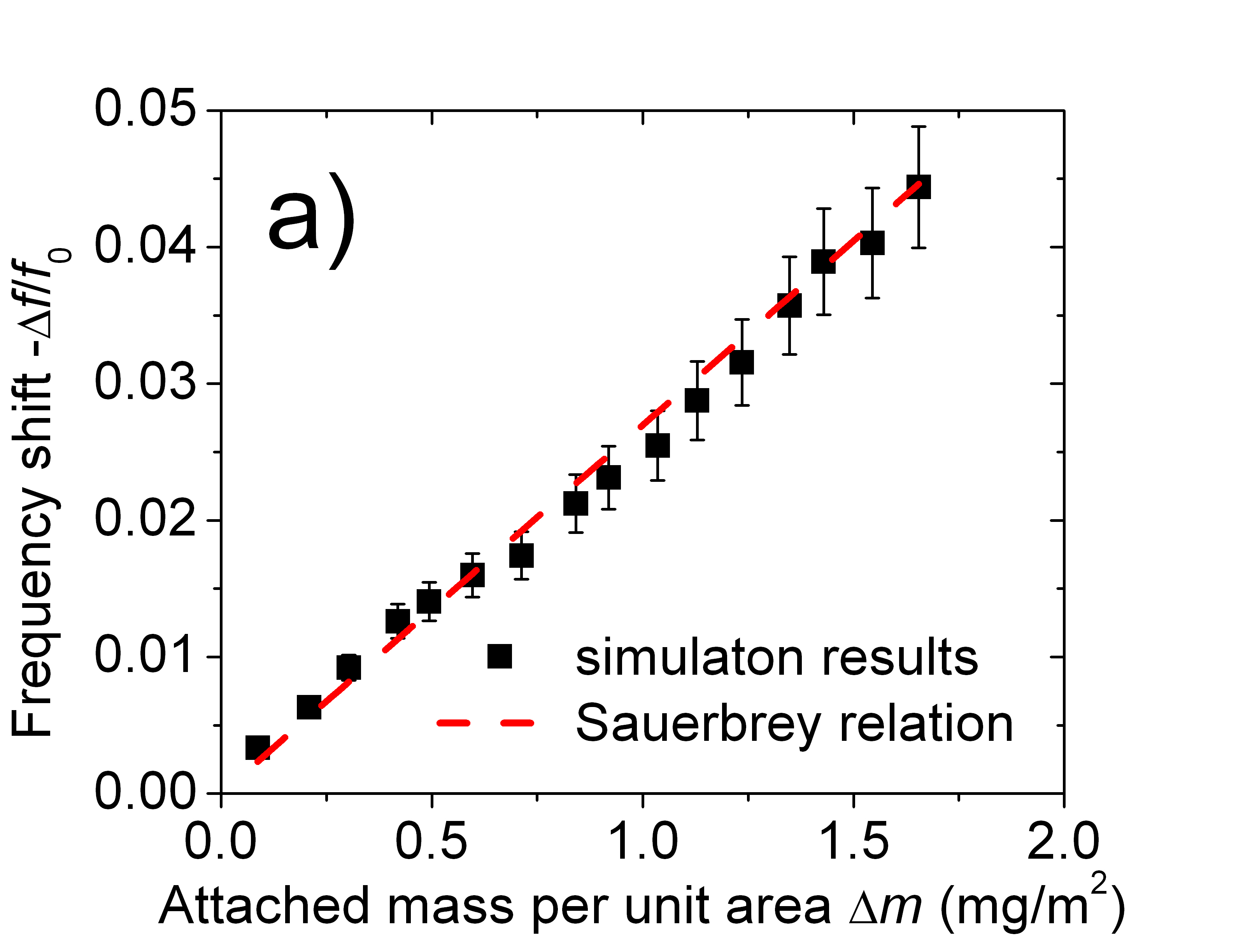} \\
\includegraphics[width=0.36\textwidth,viewport=0 0 740 540,clip=]{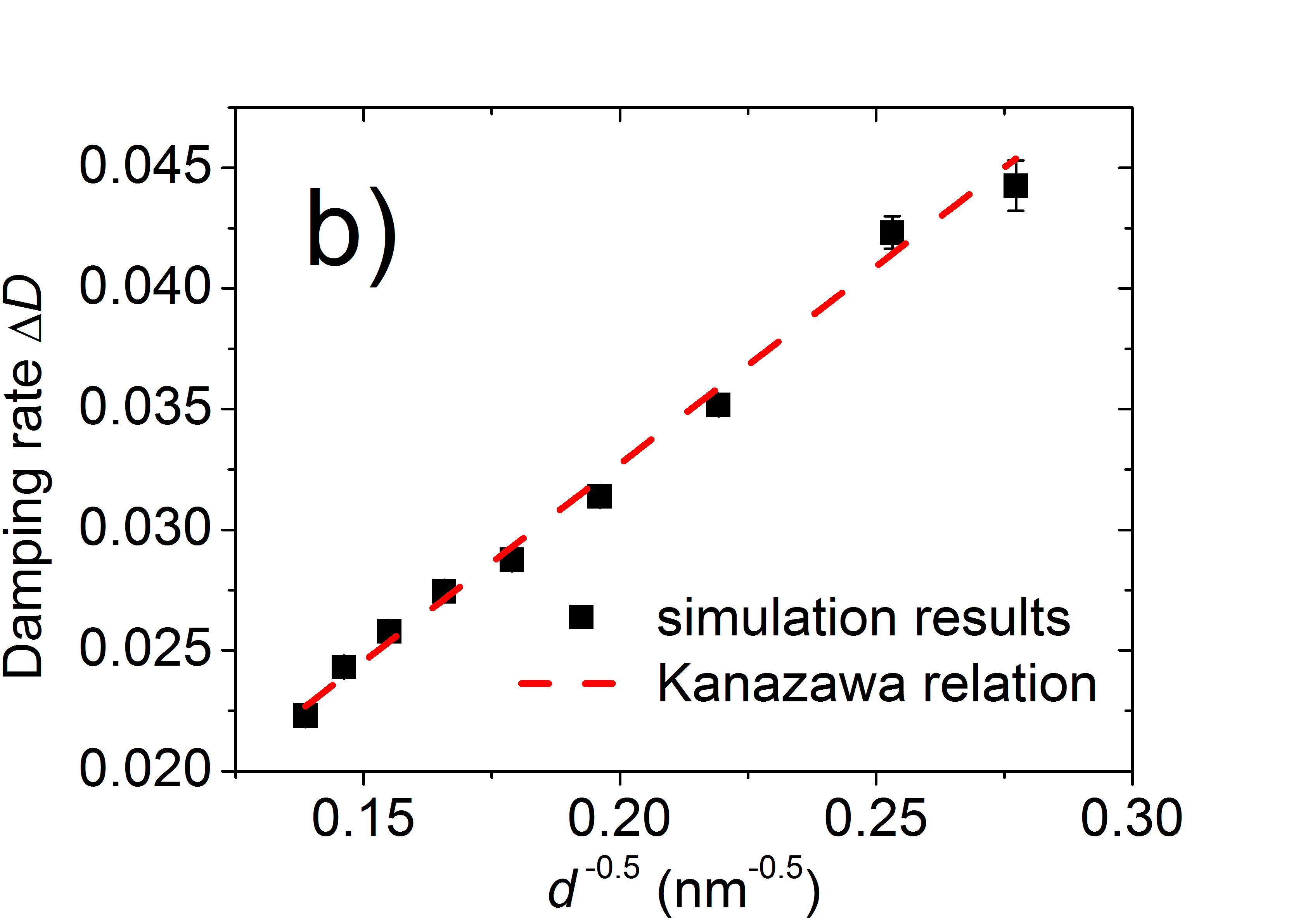}
\end{tabular}
\caption{Test of QCM model with no-slip boundary condition. a) Simulation results of rigid solid loading (squares) compared to Sauerbrey's prediction (dashed line see~\ref{e.sauerbrey}); b) Simulation results of water loading (squares) compared to Kanazawa's prediction (dashed line see~\ref{e.kanazawa_d})}
\label{f.test}
\end{figure}

\begin{figure}
\includegraphics[scale=0.25]{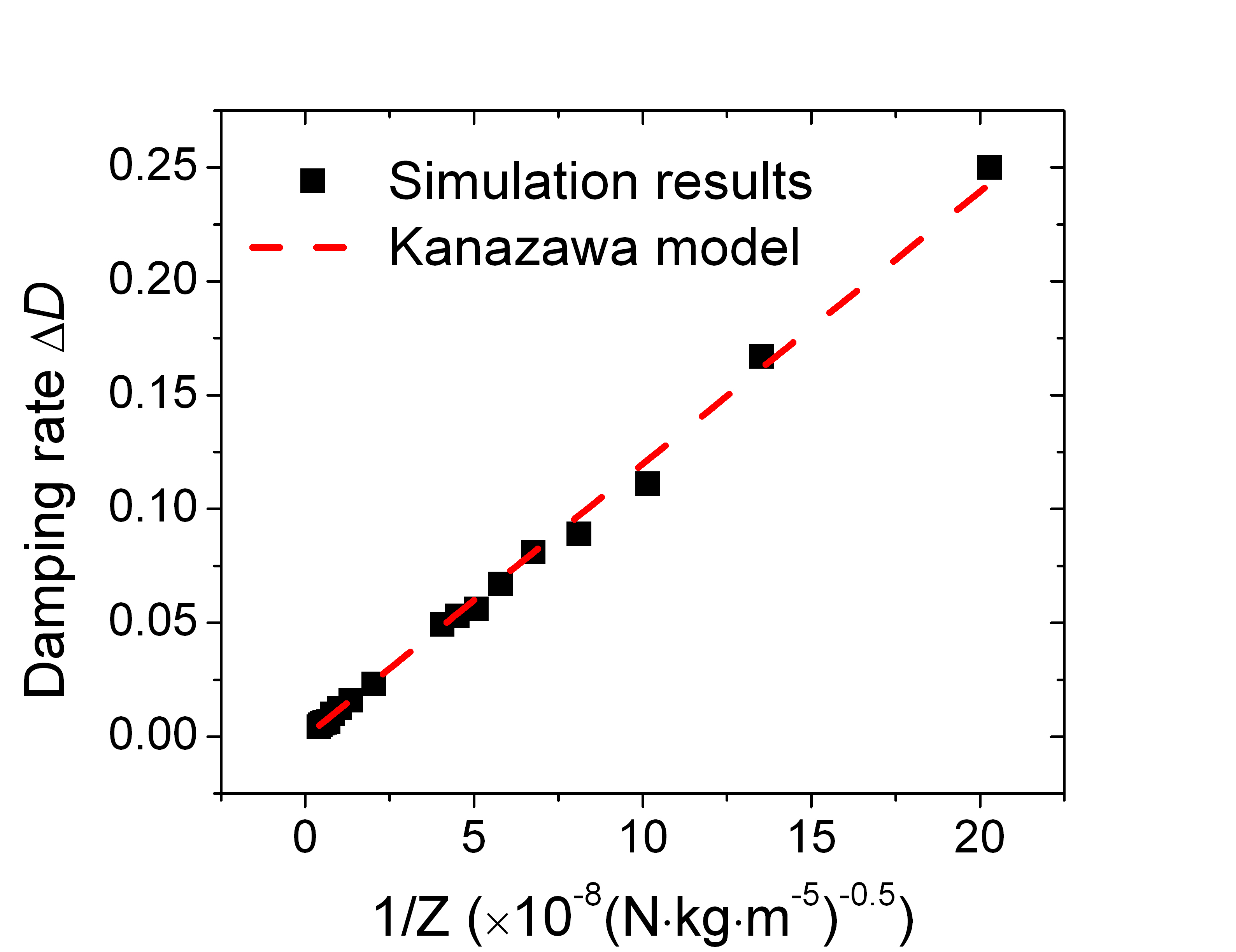}
\caption{Test of spring model with no-slip boundary condition. Simulation results of water loading (squares) compared to Kanazawa's prediction (dashed line see~\ref{e.kanazawa_d}, \(Z=\sqrt{c_{q}\rho_{q}}\))}
\label{f.spring_noslip}
\end{figure}

Mechanical response of QCM with a simple loading (e.g., rigid thin film) and a no-slip boundary condition has been studied extensively. The corresponding continuum-level theories summarized in the review section have been verified by experiments. Before investigating the effect of slip on mechanical properties of a QCM resonator, we first need to show that our MD model of QCM reproduces the correct mechanical behavior with the no-slip boundary condition. The non-loaded damping rate $D_0$ is about 0.001. Separate sets of simulations are performed for loading QCM with a rigid thin film (where we measure the resulting change $\Delta f$ in resonant frequency) and for loading QCM with water (where we measure the change $\Delta D$ in the damping rate). We control the rigid loading by attaching different numbers of atoms to the QCM surface or by modifying the atomic mass of the attachment. The two approaches yield consistent results. In the liquid loading test, we vary the crystal thickness to induce different resonant frequencies. The results of tests performed for a rigid film loading and a liquid loading are shown in Fig.~\ref{f.test}(a) and (b), respectively. An excellent agreement is found between our MD simulations and the Sauerbrey relation (Eq.~\ref{e.sauerbrey}) and Kanazawa model (Eq.~\ref{e.kanazawa_d}), for the two types of loadings, respectively. Similarly good agreement was found for the simple spring model as shown in Fig.~\ref{f.spring_noslip}. 
In summary, both the shear-wave model and the spring model capture correctly the behavior of resonators in the limit of no-slip boundary conditions. The results of the simulations agree with the Sauerbrey and Kanazawa relations for dry and aqueous conditions, respectively.

\subsection{Static slip interface}
\begin{figure}
\begin{tabular}{cc}
\includegraphics[width=0.36\textwidth,viewport=20 0 700 540,clip=]{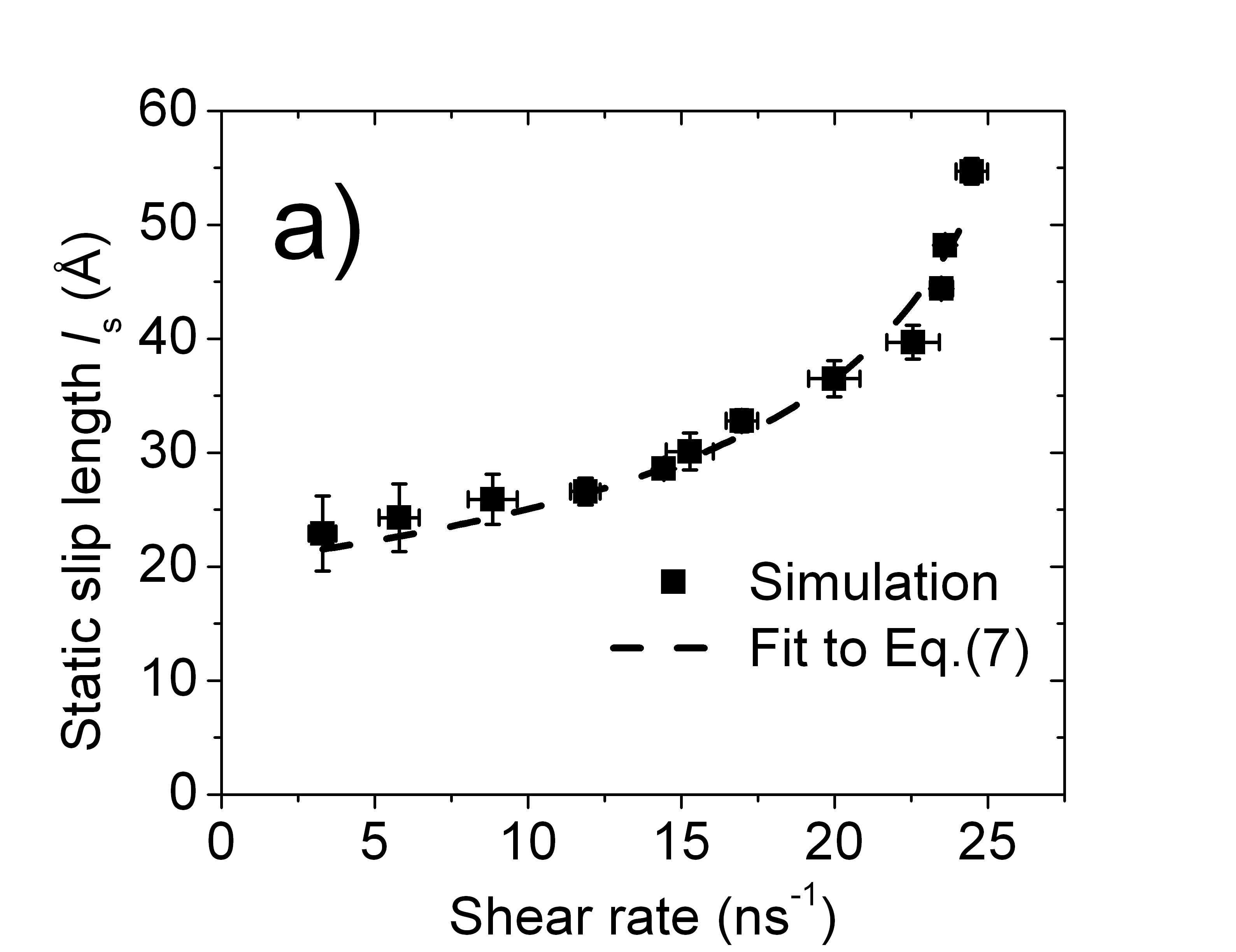} \\
\includegraphics[width=0.36\textwidth,viewport=20 0 700 540,clip=]{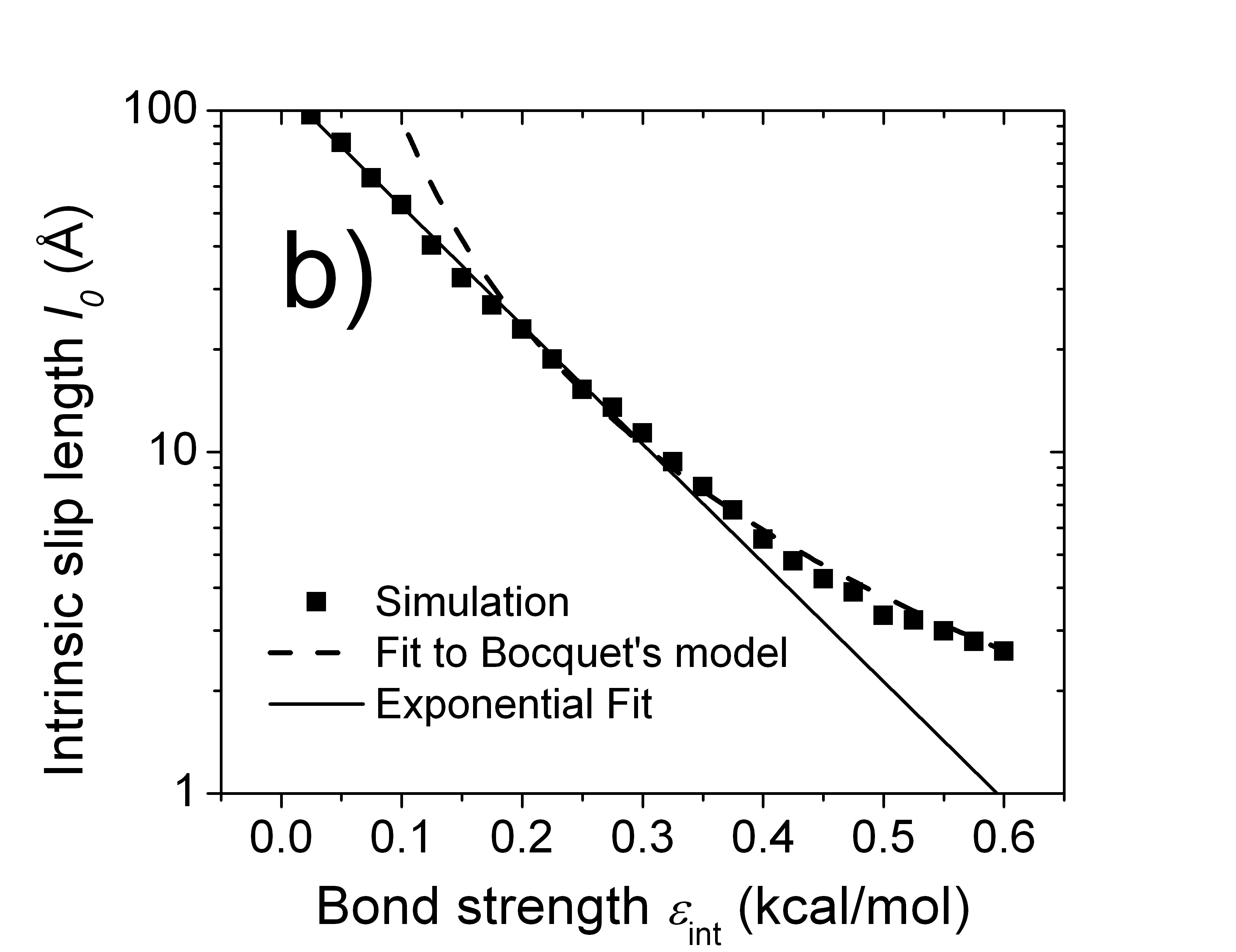}
\end{tabular}
\caption{Static slip length measured from simulations. a) Static slip length of 0.2 kcal/mol bond strength at different shear rates, fitted to \ref{e.slip_shear}. b) Static slip length as a function of bond strength. Squares correspond to the estimated intrinsic slip lengths.}
\label{f.static_slip}
\end{figure}

The next step towards the development of a model with a dynamic slip is to measure the intrinsic slip length $l_0$ defined in Eq.~\ref{e.ld}. $l_0$ does not depend on frequency and can be regarded as the dynamic slip length in the limit of zero frequency. However, because simulations with low frequencies are computationally too expensive, we measure $l_0$ using Eq.~\ref{e.ls}, that is in the static slip simulations in the limit of shear rate approaching zero. In our approach we use different sliding velocities to determine the static slip lengths $l_s$ as a function of shear rate $\dot{\gamma}$ and we estimate $l_0$ by extrapolating $l_s$ to the limit of $\dot{\gamma}\rightarrow 0$ using Eq.~\ref{e.slip_shear}.
In Fig.~\ref{f.static_slip} (a), we show the plot of $l_s$ as a function of $\dot{\gamma}$ for the case of bond strength $\varepsilon_{int}=0.2$ kcal/mol. Data obtained from MD simulations is well approximated by the empirical relationship given in Eq.~\ref{e.slip_shear}. This relationship is used to find the intrinsic slip length $l_0$ for different bond strengths $\varepsilon_{int}$, as shown in Fig.~\ref{f.static_slip} (b). We find that the dependence of $l_0$ on $\varepsilon_{int}$ is approximately exponential for $\varepsilon_{int}<0.35$ kcal/mol (solid line in Fig.~\ref{f.static_slip} (b)). The empirical relationship given by Barrat {\it et al.}~\cite{wetting-slip} (Eq.~\ref{e.slip_bond}) provides a good fit to the data for $\varepsilon_{int}>0.2$ kcal/mol (dashed line in Fig.~\ref{f.static_slip} (b)). 

\subsection{Dynamic slip interface}
\begin{figure}
\includegraphics[scale=0.05]{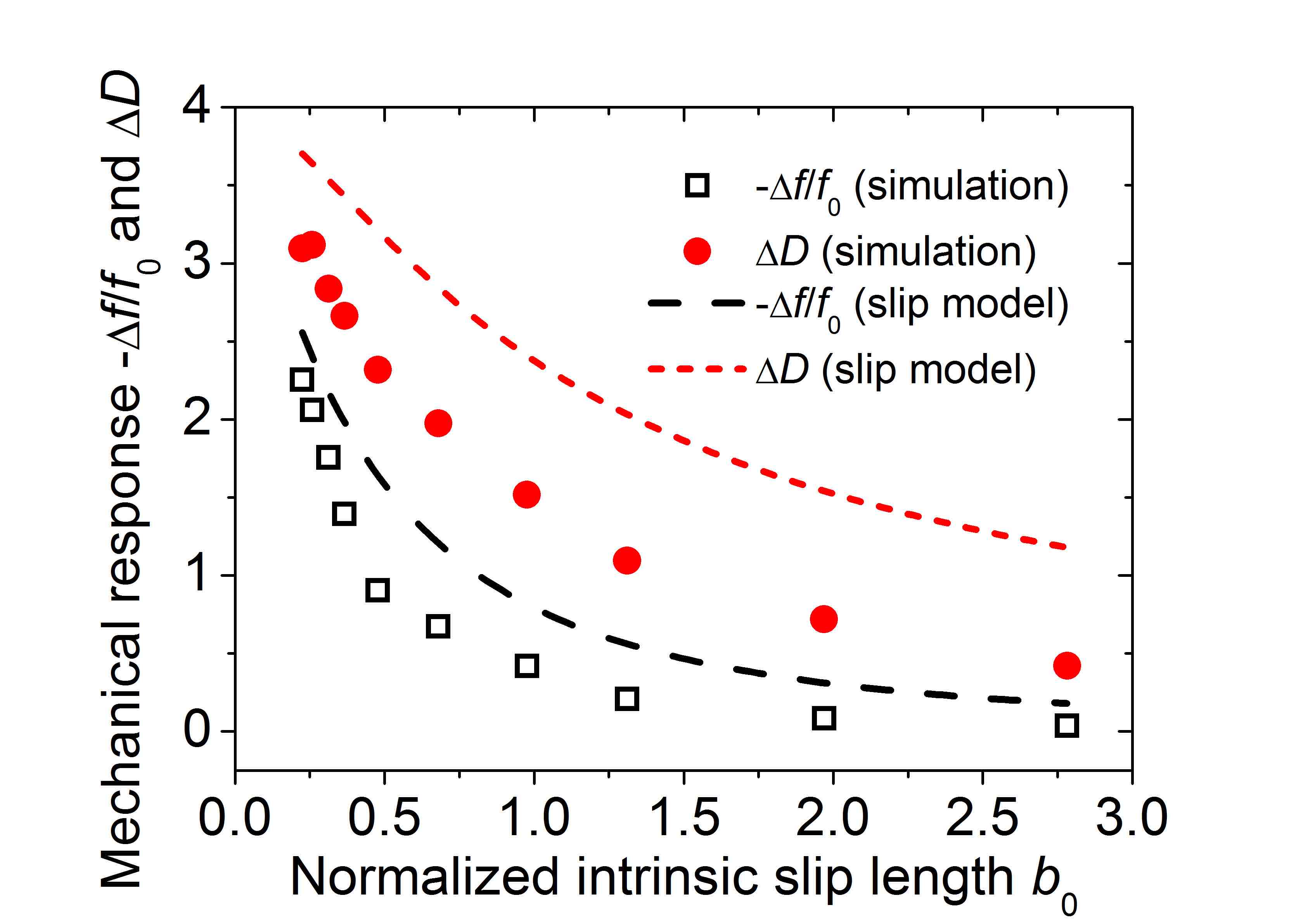}
\caption{Comparison of predictions from continuum-level slip models given by ~\ref{e.0slip_f} and~\ref{e.0slip_d} against MD measurements of mechanical response of the QCM.}
\label{f.bad_fit}
\end{figure}

As pointed out in the review section, the slip length that enters existing continuum-level slip models is treated as a single real number and no frequency dependency is considered. This treatment implicitly assumes that intrinsic slip length can be used in the dynamic friction problem on the vibrating solid/liquid interface with small amplitude, as described in Eq.~\ref{e.0slip_f} and Eq.~\ref{e.0slip_d}. We test this assumption in our MD simulations by measuring the QCM's mechanical response as a function of the intrinsic slip lengths at different frequencies, and comparing to the prediction of Eq.~\ref{e.0slip_f} and Eq.~\ref{e.0slip_d}. All simulations are performed using the spring model of QCM as the source of the resonance. As shown in Fig.~\ref{f.bad_fit}, the theoretical predictions for both frequency shifts and damping rate shifts (dashed lines) significantly overestimate the corresponding quantities measured directly in MD simulations (symbols).

Our goal here is to identify the physical phenomena that underlie the observed deviations in the dependence of frequency shift and damping rate on slip length (Fig.~\ref{f.bad_fit}) and to develop a theory that includes these phenomena. We hypothesize the following reasons for the break-down of the existing theories when applied to high-frequency resonators: a) viscosity of water depends on frequency; b) slip length depends on frequency; c) inertia of the liquid layer near the interface has a non-negligible contribution to friction force. These hypotheses are tested and discussed in the remainder of this section.

\begin{figure*}
\begin{center}$
\begin{array}{ccc}
\includegraphics[width=0.32\textwidth,viewport=10 0 720 530,clip=]{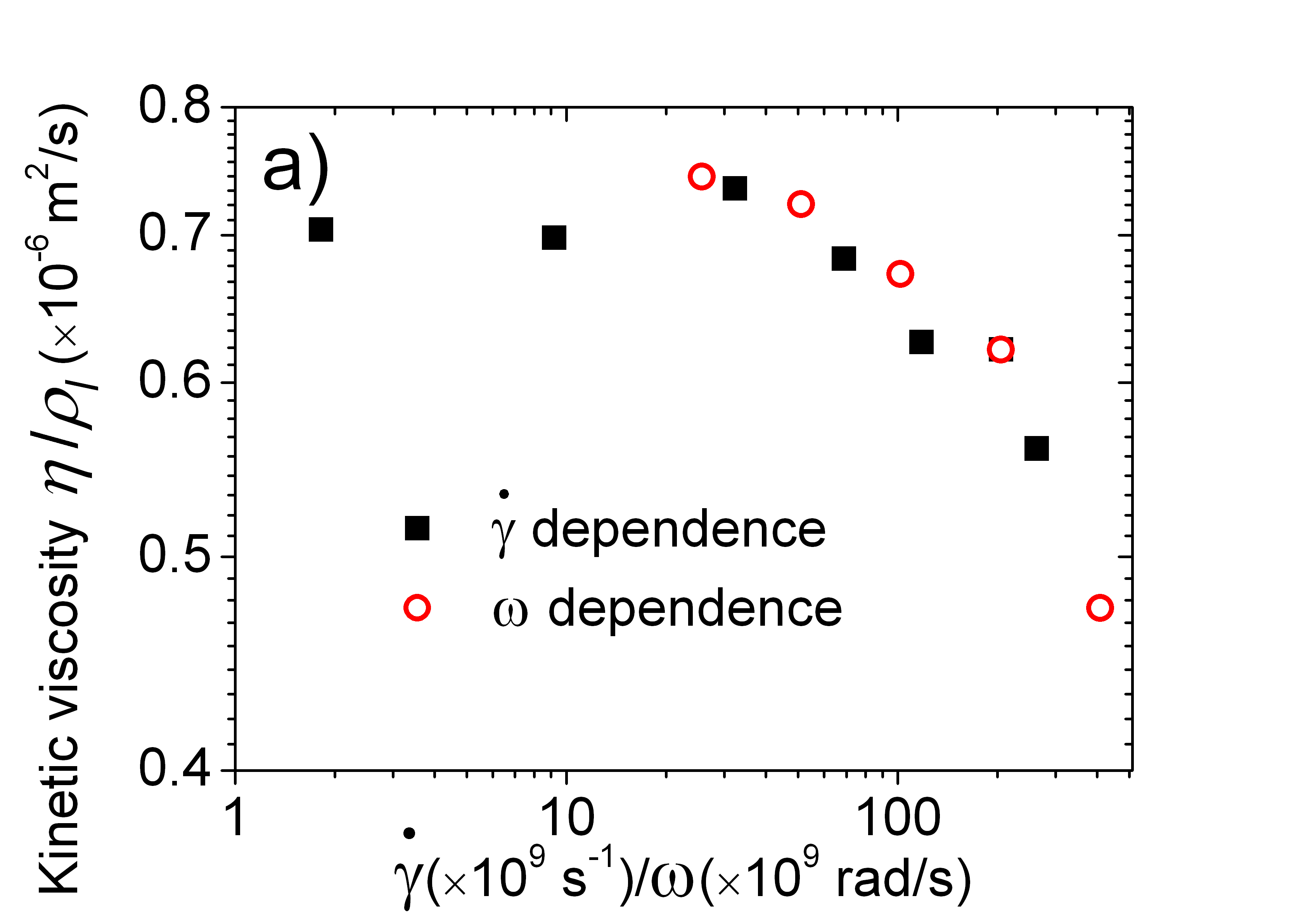} &
\includegraphics[width =0.3\textwidth,viewport=10 0 700 580,clip=]{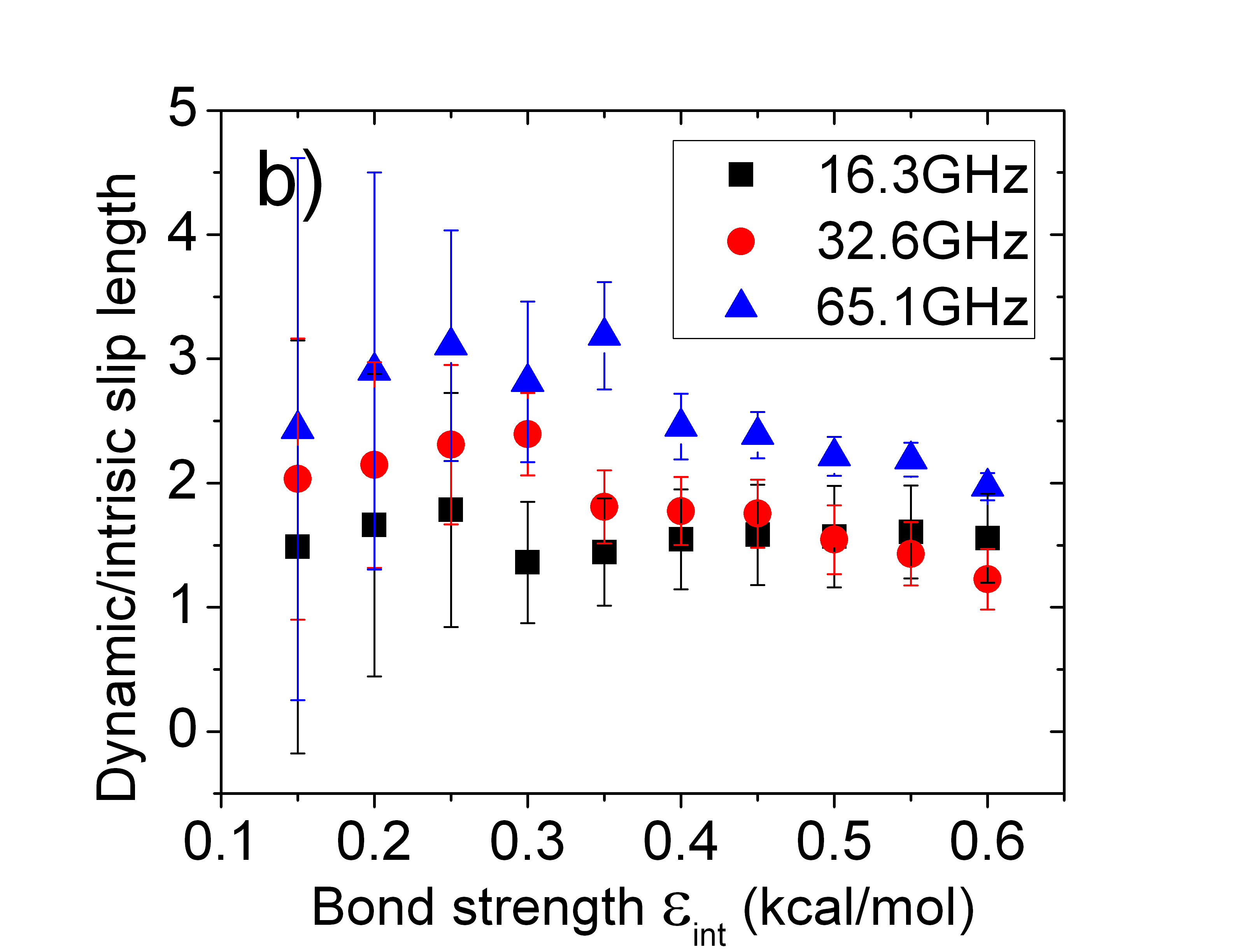} &
\includegraphics[width=0.32\textwidth,viewport=10 0 720 530,clip=]{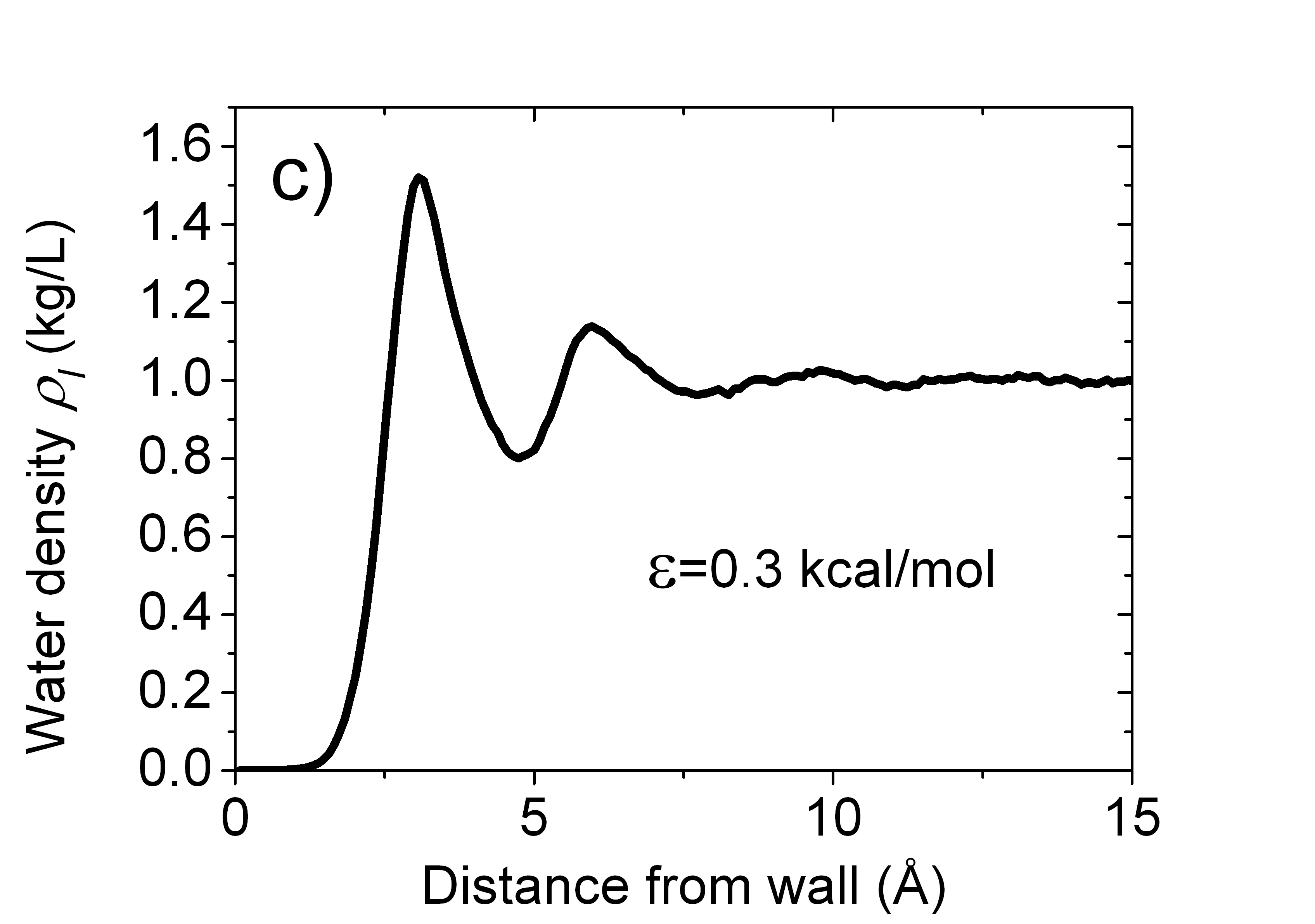}
\end{array}$
\end{center}
\caption{Effects needed to consider in a slip model. a) Frequency dependency of water viscosity; b) Frequency dependency of slip length: ratio of dynamic length and static slip length ; c) Water density profile near the interface} 
\label{f.3reasons}
\end{figure*}

We first consider the viscosity of water and determine if the assumption that the viscosity is a real constant number holds at high frequency. Based on continuum fluid mechanics, the liquid viscosity is generally dependent on shear rate and vibrational frequency. Taking the frequency dependence explicitly into account, the viscosity can be written as $\eta(\omega)=\eta'(\omega)-i\eta''(\omega)$. For a Newtonian liquid, $\eta'(\omega)\gg\eta''(\omega)$, from which it follows that the velocity profile of the damping wave can be described as~\cite{Bird}
\begin{equation}
v(z)=v_0e^{\frac{-1-i}{\delta}z},
\label{e.damp_wave}
\end{equation}
where $v$ is the shear velocity and $\delta$ is the penetration length defined in Eq.~\ref{e.penetration}. By analyzing the velocity profile of the water damping wave in our simulations, we found that Eq.~\ref{e.damp_wave} describes the amplitude and the phase of the damping wave very well. This finding implies that the imaginary part of viscosity $\eta''(\omega)$ can be ignored and we can calculate the viscosity based on the measurement of the penetration length at different frequencies, using Eq.~\ref{e.penetration} (see the supporting information). The same equation can be used to determine viscosity as a function of shear rate (if $\delta$ is measured as a function of $\dot{\gamma}$). The dependence of viscosity on both the vibrational frequency and the shear rate is shown in Fig.~\ref{f.3reasons}a. We can see that the viscosity decreases at high frequencies and/or high shear rates and it converges to $\sim7.3\times 10^{-7}$ m$^2$/s in the low frequency (or low shear rate) limit. It is interesting to point out that our data is consistent with the empirical Cox-Merz rule, which states that $\eta(\dot{\gamma})\approx |\eta(\omega)|$, if $\dot{\gamma}=\omega$. In summary, the water viscosity observed in our high frequency simulation can be treated as a real number, which decreases with increasing frequency, although not very strongly (it remains on the same order of magnitude). Because of this dependence on frequency, in our analysis we will use $\eta(\omega)$ as measured directly in our simulations instead of using the value estimated in the low-frequency limit.

The second assumption that may break down at high vibrational frequencies is that the slip length is independent of frequency. We test this assumption by calculating the ratio between the dynamic and the intrinsic slip lengths and comparing it to 1. While the intrinsic slip length $l_0$ can be directly measured in simulations (see Fig.~\ref{f.static_slip} (b)), the dynamic slip length $l_d$ cannot be measured directly. Instead, we measure the slip velocity $u_0-v_0$ and the velocity of the liquid adjacent to the solid surface $v_0$. According to Eq.~\ref{e.bcv}, the normalized dynamic slip length can be related to these two velocities as follows
\begin{equation}
b_d=\frac{|u_0-v_0|}{\sqrt{2}|v_0|}.
\label{e.dovers_slip}
\end{equation}
With this model, we can determine the normalized dynamic slip length indirectly by measuring the right hand side of Eq.~\ref{e.dovers_slip}. The limits of applicability of Eq.~\ref{e.dovers_slip} will be discussed later. If the dynamic slip length is independent of frequency, the ratio $\Gamma_d$ defined in Eq.~\ref{e.ld} should be equal to 1. In Fig.~\ref{f.3reasons} (b) we plot $\Gamma_d$ measured as a function of bond strengths for different vibrational frequencies. The ratio $\Gamma_d$ increases with increasing frequency and for the lowest frequency considered in our study (16.3 GHz) it is approximately equal to 1.5 (averaged over systems with different bond strengths). This result demonstrates that the intrinsic slip length needs to be replaced by a frequency dependent dynamic slip length to reproduce the correct physics in models of high-frequency resonators. 
Finally, we consider the effect of the inertia of the first water layer on the solid surface on mechanical properties of QCM. In particular, it is possible that the inertia of the first water layer can noticeably contribute to the momentum/energy transfer at the liquid/solid interface at high frequency. The equation of motion of the first water layer can be written as follow:
\begin{equation}
\overline{\eta}(u_0-v_{a})=-\eta\frac{\partial v}{\partial z}\Big |_{z=0}+n_{a}m\frac{\partial v_{a}}{\partial t},
\label{e.inertia}
\end{equation}
where $m$ is the mass of a single water molecule, $n_a$ is the surface number density of the first layer of water, and $v_{a}$ is the averaged velocity of first layer of water. The contribution from the first water layer to the mechanical response of the QCM is described by the second term on the right hand side of Eq.~\ref{e.inertia}. This term scales linearly with both the surface number density $n_a$ and the frequency. In our static sliding system, as there is no acceleration, the second term on the right hand side of Eq.~\ref{e.inertia} is rigorously equal to zero. In order to determine $n_a$, we count the number of water molecules in the first layer, where the extent of this layer is determined from a density profile shown in Fig.~\ref{f.3reasons} (c). To make the units consistent with the slip length and the penetration length, we introduce an inertia length $l_{a}$ and a normalized inertia length, which are, respectively, defined as 
\begin{equation}
l_{a}=n_a/n,
\label{e.adsorption_length}
\end{equation}
\begin{equation}
a(\omega)=l_{a}/\delta(\omega).
\end{equation}
In the above equations, $n$ is the number density of bulk water molecules and by writing $a(\omega)$ we explicitly indicate that inertia length depends on frequency.
We find from simulations that $l_{a}\sim 4$\AA~and that this value is not sensitive to the bond strength. Specifically, in our simulations changing the bond strength does not affect the position of the first peak in water density profile, but it affects the height and width of the peak (more hydrophilic surfaces have a higher and a narrower peak). The velocity amplitude $|v_{a}|$ of the first water layer is expected to be smaller than the velocity amplitude $|u_0|$ of the solid and larger than the velocity amplitude $|v_0|$ of the next water layer in the bulk liquid. As we did not observe any jump in the shear velocity of water, we assume $v_{a} \approx v_0$. Although this assumption is not as intuitive as $v_{a}\approx u_0$, which means a rigid adsorption, it is applicable for a wider range of interfacial bond strengths. For strong bonding (hydrophilic surfaces), $v_{a}\approx v_0\approx u_0$, since the slip length and therefore the slip velocity ($u_0-v_0$) are small. However, for weaker bonding (hydrophobic surfaces), the liquid-liquid attraction is larger than the liquid-solid attraction, making $v_{a}$ closer to $v_0$. Therefore, $v_{a}\approx v_0$ is a good approximation for all the bond strengths. The inertia term in Eq.~\ref{e.inertia} can be regarded as an additional friction force on the surface, whose contributions to the total friction scales with $a$. The value of $a$ increases with frequency. The largest $a$ found in our simulations is about 0.27, and it cannot be ignored in the analysis. We have modified the relationship between the dynamic slip velocity and the slip velocity (Eq.~\ref{e.dovers_slip}) to include the inertia effect of the first water layer. The new relationship has the following form (details in supporting information)
\begin{equation}
b_d=\frac{1}{\sqrt{2+4a+4a^2}}\Big |\frac{u_0-v_0}{v_0}\Big |.
\label{e.dovers_slip_a}
\end{equation}
We will later use Eq.~\ref{e.dovers_slip_a} to estimate the dynamic slip length in our simulations.

In summary, we found three physical phenomena that have not been included in existing models for mechanical dissipation of resonators and that are important at high frequencies. These are: viscosity of liquid depends on frequency, slip length depends on frequency, and inertia of the first layer of liquid contributes to the friction force. We now propose a new model that takes these phenomena into account.
We begin by writing the equation of motion for a spring-model of a solid vibrating along the x direction
\begin{equation}
M\omega^2x=M\omega_0^2x+F_{f}, 
\end{equation}
where $F_{f}$ is the friction force (equal to the left hand side of Eq.~\ref{e.inertia}), $x$ is the displacement of the solid, and $M$ is the mass of the solid. $\omega_0$ is the resonance frequency without friction and $\omega$ is the new frequency with friction. By solving the above equation for $\omega$ and defining $\Delta\omega=\omega-\omega_0$, we obtain
\begin{equation}
\Delta\omega=\omega-\omega_0=\frac{F_{f}}{2M\omega_0x}. 
\end{equation}
The frequency shift $\Delta f$ and the damping rate shift $\Delta D$ can be calculated as the real and imaginary part of $\Delta\omega$, that is
\begin{equation}
\Delta f=\frac{\mathrm{Im}(\Delta\omega)}{2\pi},
\end{equation}
\begin{equation}
\Delta D=\frac{\mathrm{Re}(\Delta\omega)}{f_0}.
\end{equation}
A slip boundary condition Eq.~\ref{e.inertia}, which takes into account the inertia of the first liquid layer, is then used to complete the set of equations for the new slip model and to make the equations solvable. One can derive (details in supporting information) the following relations between the mechanical response of QCM (frequency and damping shifts) and the normalized dynamic slip length $b_d$:
\begin{equation}
\frac{\Delta f}{f_0}=-\frac{1}{2Z}\sqrt{\frac{\rho_{l}\eta_d\omega}{2}}\frac{1+2a}{1+2b_d+[(1+2a)^2+1]b_d^2},
\label{e.f_new}
\end{equation}
\begin{equation}
\frac{\Delta D}{2\pi}=\frac{1}{2Z}\sqrt{\frac{\rho_{l}\eta_d\omega}{2}}\frac{1+[(1+2a)^2+1]b_d}{1+2b_d+[(1+2a)^2+1]b_d^2}.
\label{e.d_new}
\end{equation} 

\begin{figure*}
\begin{tabular}{cc}
\includegraphics[width=0.43\textwidth,viewport=15 0 710 530,clip=]{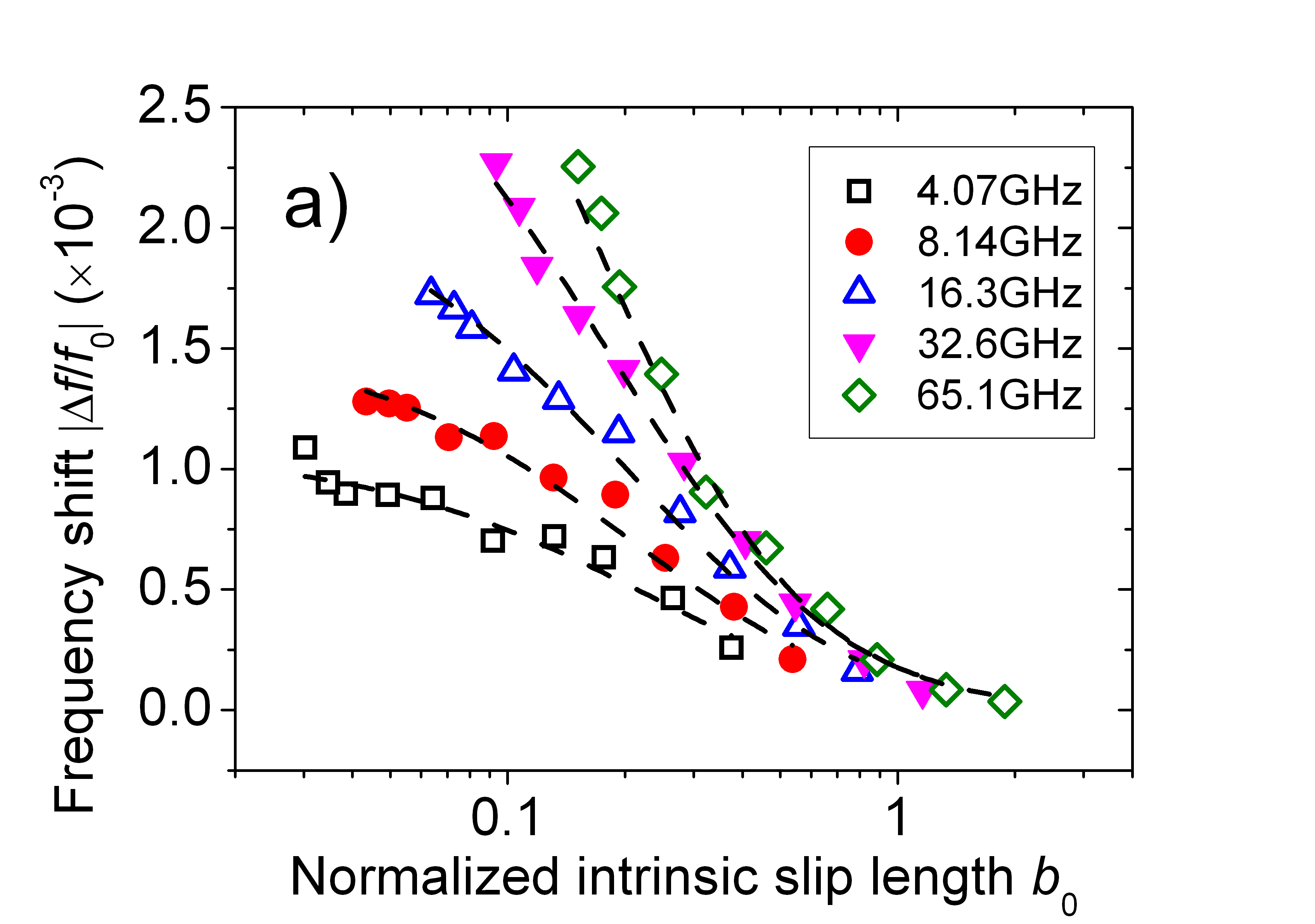} & 
\includegraphics[width=0.43\textwidth,viewport=15 0 710 530,clip=]{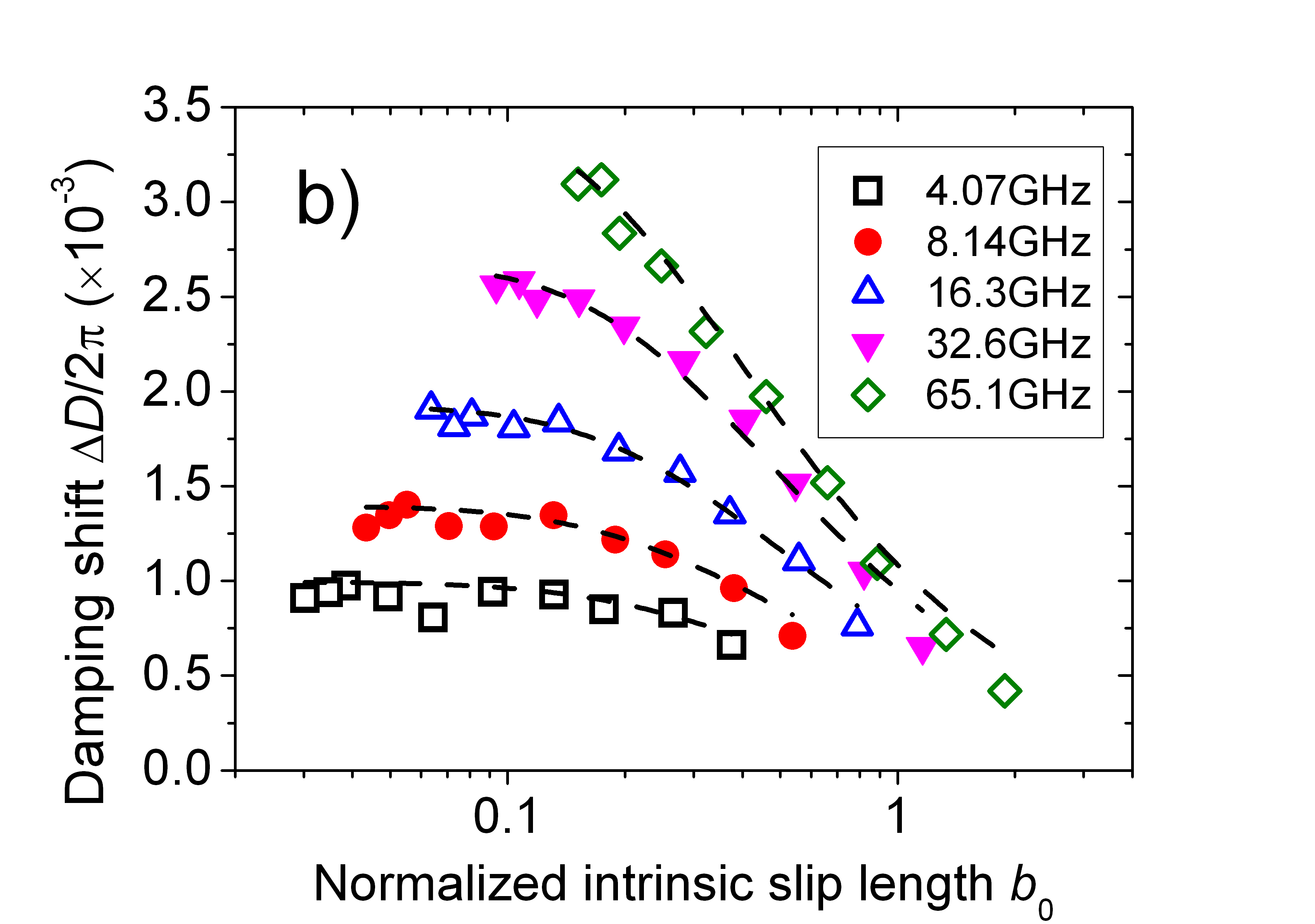} 
\end{tabular}
\caption{Results of MD simulations (symbols) for a) frequency shift and b) damping shift as a function of reduced slip length $b_0$. Dashed lines represent the predictions from our analytical model described in the text.}
\label{f.main}
\end{figure*}

\begin{figure}
\includegraphics[scale =0.25,viewport=20 0 720 530]{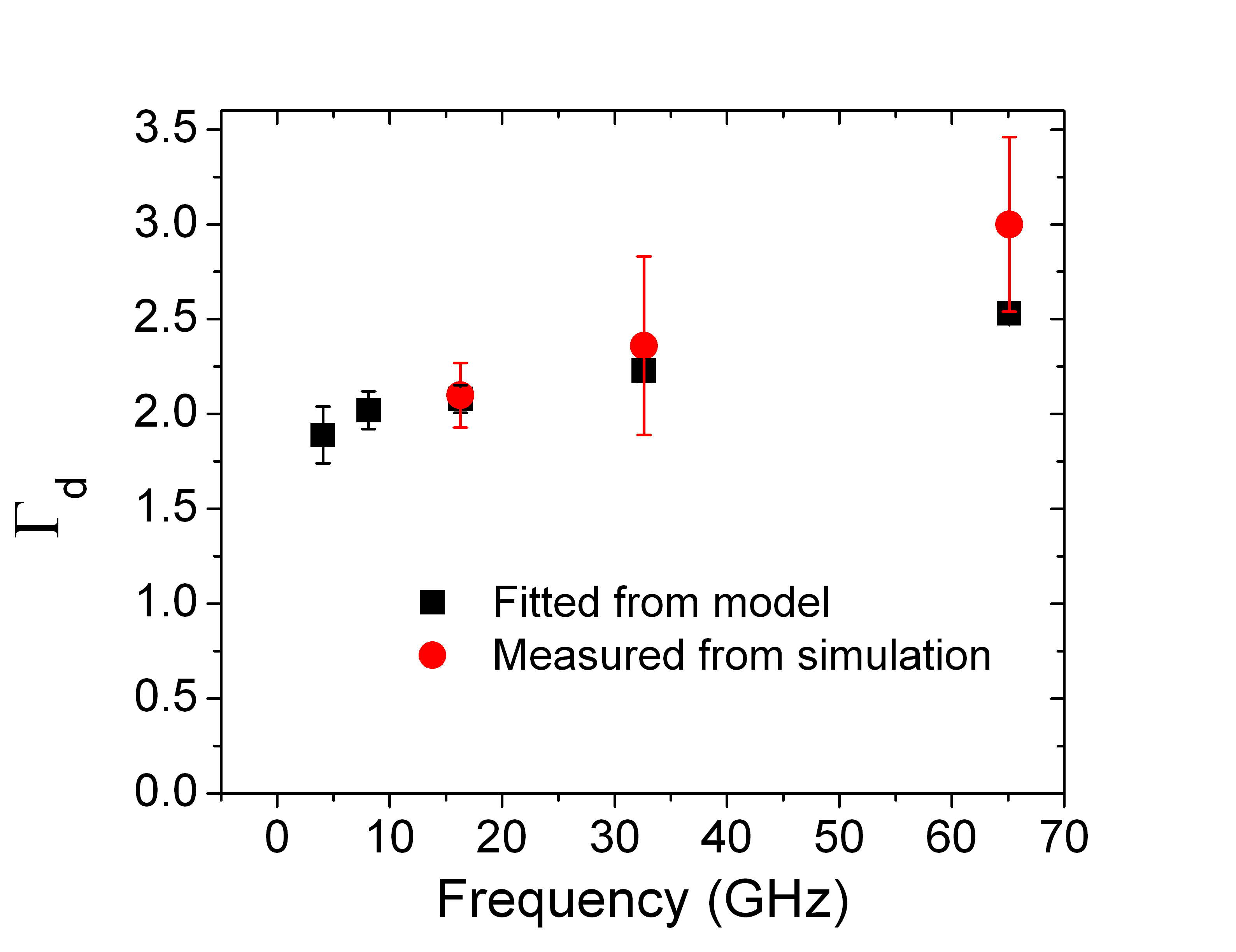}
\caption{Dependence of the ratio between dynamic and intrinsic slip length on frequency.}
\label{f.scaling_factor}
\end{figure}

To test applicability of the new model, in Fig.~\ref{f.main} we plot the frequency response $\left|\frac{\Delta f}{f_0}\right|$ (or $-\frac{\Delta f}{f_0}$ as frequency always decreases in our case) and the damping response $\frac{\Delta D}{2\pi}$ as a function of the normalized slip length for five different frequencies. These plots demonstrate that slip has dramatic consequences on both frequency shift and energy dissipation in our high frequency resonator, especially when the slip length is comparable to the penetration length of liquid. Both the absolute value of frequency shift and the damping rate shift decrease with slip length. In Fig.~\ref{f.main} at the same normalized slip length, the absolute values of frequency shift and damping rate shift are generally larger when the resonant frequency is higher. Both the trends and the quantitative data obtained in MD simulations (symbols in Fig.~\ref{f.main}) are well described by our model (dashed lines). In order to predict the frequency shift and damping rate shift using our model, we need to know the surface density of the first water layer or the inertia length $l_a$. A simple estimation of the surface number density $n_a$ and inertia length $l_a$ in Eq.~\ref{e.adsorption_length} is $n_a$=$n^{\frac{2}{3}}$ and $l_a$=$n^{-\frac{1}{3}}$=3.1~\AA, where $n$ is the number density of bulk water. The exact values of $n_a$ and $l_a$ depend on chemistry and structure of the interface. We found, however, that this dependence is not strong and in our simulations, an inertia length of 3.8$\pm$0.3~\AA~fits well the mechanical response of QCM ($\frac{\Delta f}{f_0}$ and $\Delta D$) at all frequencies and bond strengths. 
The reader should note that in Fig.~\ref{f.main} we plotted the results against the intrinsic slip length $l_0$, since it is a physical quantity that is typically measured in slip experiments and often studied in computer simulations. Our data can be easily converted to be a function of the dynamic slip length using the scaling factor $\Gamma_d$, as defined in Eq.~\ref{e.ld}. This factor represents the ratio of the dynamic and the static slip lengths. $\Gamma_d$ can be either determined by measuring the dynamic slip length using Eq.~\ref{e.dovers_slip_a} (as was done in our simulations) or by fitting the measured mechanical response ($\Delta f$ or $\Delta D$) to the equations of our model. We plot the values of $\Gamma_d$ obtained using the two methods as a function of frequency in Fig.~\ref{f.scaling_factor} and we find a good agreement between the estimates within the error bar of calculations. 

\begin{figure*}
\begin{tabular}{cc}
\includegraphics[width=0.44\textwidth,viewport=0 0 680 540,clip=]{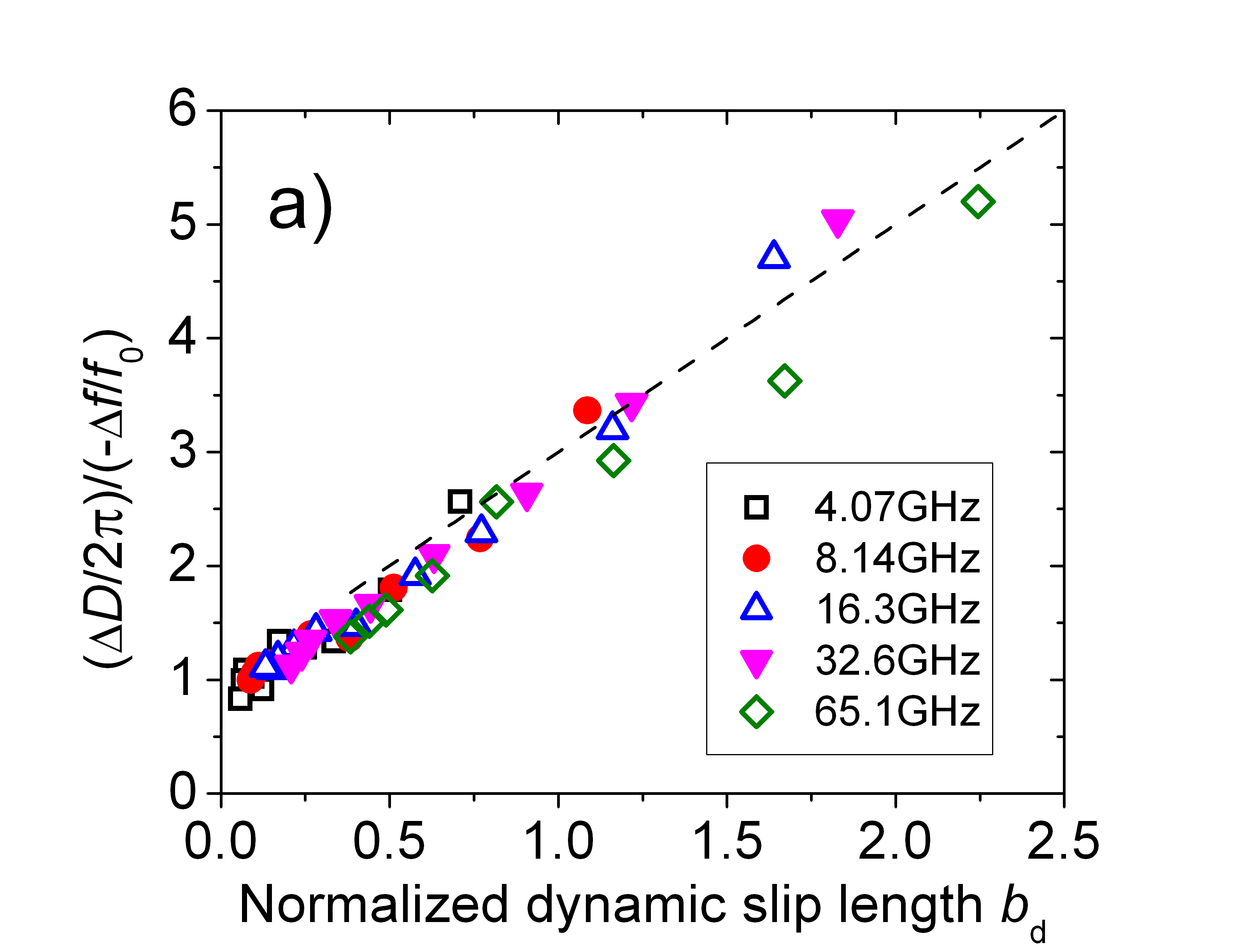} &
\includegraphics[ width=0.44\textwidth,viewport=0 0 680 540,clip=]{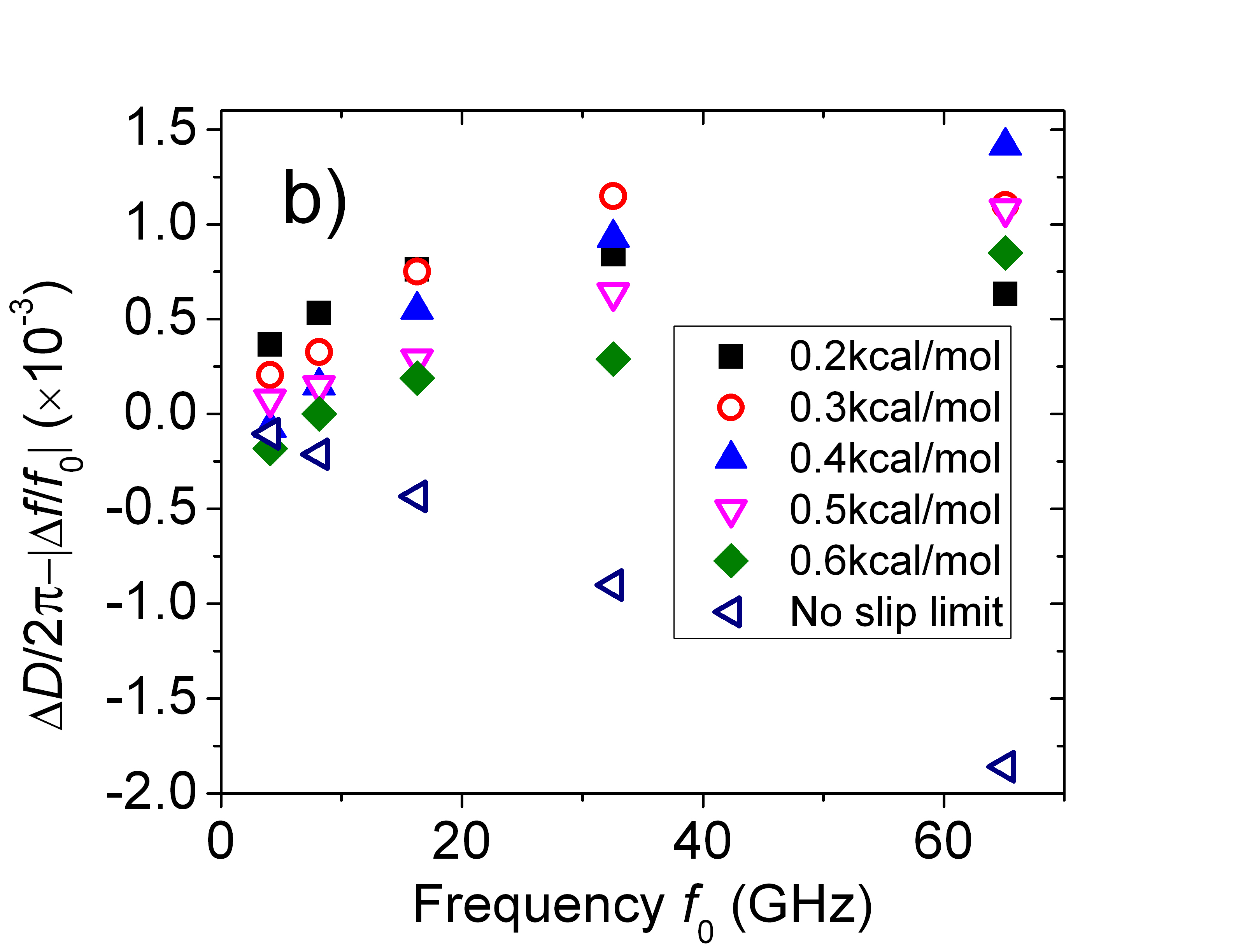}
\end{tabular} 
\caption{Comparison between frequency shift and damping rate shift. a) The ratio between damping rate shift and frequency shift (symbols) as a function of the reduced dynamic slip length. (dash line is a universal linear relationship \(y=2x+1\)). b) The sum of damping rate shift and frequency shift (or the difference of their magnitudes) plotted as a function of frequency for different interfacial bond strength $\varepsilon_{int}$.}
\label{f.dandf}
\end{figure*}

As will be demonstrated below, it is useful to rewrite Eq.~\ref{e.f_new} and Eq.~\ref{e.d_new} as the ratio and the difference of the damping shift $\frac{\Delta D}{2\pi}$ and the absolute value of frequency shift $\left|\frac{\Delta f}{f_0}\right|$, namely,
\begin{equation}
\frac{\Delta D/2\pi}{|\Delta f/f_0|}=\frac{1+[(1+2a)^2+1]b_d}{1+2a},
\label{e.d/f}
\end{equation}
\begin{equation}
\frac{\Delta D}{2\pi}-\left|\frac{\Delta f}{f_0}\right|=\frac{1}{\pi Z}\sqrt{\frac{\rho_{l}\eta_d\omega}{2}}\frac{[(1+2a)^2+1]b_d-2a}{1+2b_d+[(1+2a)^2+1]b_d^2}.
\label{e.d+f}
\end{equation}
In Fig.~\ref{f.dandf} (a) we plot the ratio in Fig.~\ref{e.d/f} as a function of $b_d$ for data calculated from MD simulations. Irrespectively of the frequency, all simulation data falls on the same line. This linear dependence can be understood by considering that $a$ is usually smaller than 1 (the largest $a$ in our simulation is about 0.27), which means that $l_a < \delta$. With that in mind, we can simplify the expression for the ratio between damping rate and frequency shift (Fig.~\ref{e.d/f}) to the $1^{\mathrm{st}}$ order of $a$ as follows
\begin{equation}
\frac{\Delta D/2\pi}{|\Delta f/f_0|}\approx 2b_d+1-2a.
\label{e.linear}
\end{equation}
The above expression can be furthermore simplified to the $0^{\mathrm{th}}$ order of $a$ and the right hand side of Eq.~\ref{e.linear} is approximately equal to $2b_d+1$. This relationship is plotted as a dashed line in Fig.~\ref{f.dandf} (a) and it shows a good agreement with the MD data. The one-to-one correspondence between the normalized dynamic slip length and the ratio in Eq.~\ref{e.d/f} provides an easy way of estimating the slip length from QCM measurements. This estimation is generally more accurate when the ratio is large so that the contributions to the ratio from any source (e.g., interfacial inertia) other than slip can be neglected. In other words, if the normalized slip length is too small (as compared to 1), one cannot determine its value from Eq.~\ref{e.linear}. 
For small slip lengths, the relationship $\frac{\Delta D/2\pi}{|\Delta f/f_0|}\approx 2b_d+1$ is not a good approximation. According to our model, in this limit it is possible to obtain some qualitative information about the slip length and more specifically one can determine if the normalized slip length $b_d$ is smaller, larger, or comparable to the normalized inertia length $a$, where the latter quantity is a measure of the width of the interface. This comparison can be accomplished by analyzing the difference $\frac{\Delta D}{2\pi}$ and $\left|\frac{\Delta f}{f_0}\right|$, because the sign of the expression in Eq.~\ref{e.d+f} is well approximated by the sign of $b_d-a$, that is
\begin{equation}
sign\left[\frac{\Delta D}{2\pi}-\left|\frac{\Delta f}{f_0}\right|\right]=sign\left\{[(1+2a)^2+1]b_d-2a\right\}\approx sign(b_d-a),
\label{e.sign}
\end{equation}
which means when the normalized dynamic slip length $b_d$ is smaller than the normalized inertia length $a$, the frequency shift is larger than the damping rate shift and vice versa. A negative value of the difference in Eq.~\ref{e.d+f} is not predicted by either the no-slip model (Kanazawa model~\cite{kanazawa}) or by earlier slip models~\cite{Hayward,Daikhin, Mchale, Zhuang, Persson} that ignore the inertia of the interfacial liquid. We plot the difference in Eq.~\ref{e.d+f} in Eq.~\ref{f.dandf} (b) as a function of frequency for various bond strengths (slip lengths). As shown by our MD data (symbols), the expression given by Eq.~\ref{e.d+f} can become negative in the no-slip limit or when the normalized slip length is smaller than normalized inertia length. This finding from simulation is again consistent with our analytical model (Eq.~\ref{e.sign}) and provides a possible explanation to the origin of the negative difference in Eq.~\ref{e.d+f} observed in some QCM experiments~\cite{Mchale}.
Additionally, from the curves in Fig.~\ref{f.dandf} (b) we can see that independently of whether the difference is positive or negative, the absolute value of this difference generally increases with increasing frequency, which is consistent with trends observed in experiments~\cite{bubble-slip}.

\section{Discussion and conclusion}
It is yet instructive to discuss possible limits of applicability of our model and under what conditions this model becomes necessary and outperforms earlier low-frequency models. First of all, although the model has been tested against simulations performed at high frequencies, it is expected to apply also in the limit of low frequencies. It is because there is no discontinuous change in viscosity, slip length, and interfacial inertia as a function of frequency and the dependence of these quantities on frequency is monotonic (see Fig.~\ref{f.3reasons}). Contributions from the three phenomena identified in this paper (viscosity and slip length dependence on frequency and interfacial inertia) are present at low frequencies, but these contributions will be negligible in the zero frequency limit. In fact, as shown in supporting information, in the zero frequency limit our generalized model will be reduced to the previously developed models summarized in Eq.~\ref{e.0slip_f} and Eq.~\ref{e.0slip_d} and therefore our model can be thought of as a generalized approach.
Frequency enters into the slip model in many different ways. First of all, penetration length that characterizes the dimensions of the liquid wave is dependent on frequency. From Eq.~\ref{e.0slip_f} and Eq.~\ref{e.0slip_d} one can see that it is the normalized slip length that governs the change of frequency shift and damping shift. The smaller the normalization factor (penetration length), the bigger the normalized slip length. Since the slip length is typically in the nm regime, models assuming no-slip boundary condition that work very well for macroscopic systems will begin to fail with the penetration length being reduced to the nm length scales. Taking QCM as an example, a fundamental frequency of about 5 MHz will lead to a penetration length of about 250 nm for water. Assuming the slip length is 10 nm, slip will cause a decrease of 7.7\% in the absolute value of the frequency shift and 0.3\% in the damping rate shift, as compared to the no-slip condition. Consequently, ignoring the slip will lead to 7.7\% and 0.3\% errors in frequency shift and damping rate shift, respectively, which effect is not dramatic. However, if the operating frequency of QCM is increased to 500 MHz (which corresponds to a 25 nm penetration length for water), the same amount of slip will result in 53\% and 15\% errors in frequency shift and damping rate shift, respectively. In this case, it is necessary to use a slip model to predict mechanical behavior of QCM. 
As shown by our simulations on water, when the frequency is as high as a few hundred MHz or higher, the slip length may be quite different from that measured in the static shearing experiments or simulations. This is a somewhat surprising phenomenon that has not been previously reported in literature. We expect this phenomenon to occur in a broad range of liquids, since most liquids have a longer relaxation time than water. Thus our results suggest that in typical QCM experiments with polymeric liquids, one should use the concept of a dynamic slip length and a generalized slip model that considers frequency effects. The identified frequency dependence of slip length also suggests that the liquid/solid friction coefficient may need to be treated as interfacial viscosity. Mechanical analog models, similar to those already developed for liquid viscosity, may be useful in describing solid/liquid friction and in identifying underlying physics. In fact, simple mechanical analog models of solid/liquid friction have been already proposed to shed light on certain experimental observations~\cite{complex-slip,voigt}. On the other hand, molecular-level understanding of frequency dependence of slip length is still missing and providing such understanding is beyond the scope of this paper.
Another phenomenon that enters our generalized model is the inertia of the first liquid layer. One should be aware of the difference between this interfacial inertia and the adsorption on surface, although both these effects lead to an increase in the magnitude of frequency shift. Adsorption requires a much stronger interaction between the solid and the liquid molecules and if adsorption takes place, the liquid density profile near the solid surface is expected to have a much sharper peak than that observed in our simulations (see Fig.~\ref{f.3reasons} (c)). Velocity of the adsorbed layer should be equal to the velocity of the solid wall and consequently the slip can only take place between the adsorbed liquid layer and the liquid above it. For atomically smooth surfaces, slip between the adsorbed layer and the liquid is not likely to happen. The inertia effect from the interfacial liquid is more general than adsorption, as it is not limited to the case of strong interactions between liquid and solid. Our treatment on the first layer water is a simple way to include effects from the interfacial region, where the properties of liquid, such as density and viscosity, differ from those in the bulk liquid. This approach is more accurate than the sharp interface condition that assumes the width of interfacial region to be zero. For water on our atomically smooth surface, we found the interfacial layer to be about one monolayer thick, however this thickness may vary depending on the surface conditions. In general, the width of the interfacial region is expected to be on the order of a few molecular diameters. 
Since in currently used QCM technology, typical surfaces are not atomically smooth, it is interesting to ask about the effects of surface roughness on solid/liquid friction. This topic is an active area of research~\cite{interfacial-water,roughness-induced-slip,trapped-water,roughness,Mchale} and many insights have been brought through MD simulations~\cite{interfacial-water}. In most cases, roughness was shown to reduce the slip length. Roughness have been also postulated to be responsible for the negative value of the difference between the damping rate shift and the magnitude of the frequency shift (Eq.~\ref{e.d+f}) observed in in some QCM experiments~\cite{bubble-slip}. This negative value is not predicted by previous slip-boundary models. To explain this phenomenon, McHale {\it et al.}~\cite{Mchale} introduced the concept of a negative slip length, which was assumed to be the consequence of surface roughness. Our model provides a possible alternative explanation of the experimental observation without the need to invoke a negative slip length. It is likely that roughness affects both the slip length and the inertia length (or the width of the interfacial region). A reduced (although still positive) slip length and an increased inertia length in our model will result in a negative value of the difference given by Eq.~\ref{e.d+f} and therefore this model may explain the experimentally observed effects of roughness. The effects of roughness are expected to be much less important when the size of shear-wave acoustic resonators is reduced, such as in the case of MEMS and NEMS devices.

In summary, the effects of slip boundary condition have been investigated by MD simulation. We discovered new phenomena that emerge at high vibrational frequencies. For example, we have shown that slip length is frequency dependent and to account for this dependence explicitly, we introduce a concept of a dynamic slip length. We have also shown that the interface between solid and liquid cannot be treated as sharp and that the inertia of water near the interface contributes to friction. A generalized slip model that includes newly discovered high-frequency phenomena is developed to connect the slip length and mechanical response of high frequency acoustic resonators. The model shows excellent agreement with MD simulations. A linear relationship between the ratio of measured mechanical properties and the slip length is discovered. This relationship provides a means for determining slip length experimentally, which had been an outstanding challenge in the field.
\begin{acknowledgement}
The authors gratefully acknowledge support from the NSF grant CMMI-0747661.
\end{acknowledgement}

\bibliography{GK reference}
\end{document}